\documentclass[12pt]{article}
\usepackage{amsmath,amssymb}
\usepackage[dvips]{graphicx}
\usepackage{array,booktabs}
\begin{document}
\title{ Anomalous radiative transitions}
\author{{Kenzo Ishikawa}${}^{1}$, {Toshiki Tajima}${}^{2,3}$, {and} 
{Yutaka Tobita}${}^{1}$} 
\maketitle
\begin{center}
 ${}^{1}$ Department of Physics, Faculty of Science, 
Hokkaido University, Sapporo 060-0810, Japan \\
${}^{2}$ Department of Physics and Astronomy, University of California,
Irvine, CA 92697, USA\\
${}^{3}$ KEK High Energy Accelerator Research
 Organization, Tsukuba 305-0801, Japan
\end{center}


\begin{abstract}
Anomalous transitions involving  photons derived by  many-body
 interaction  of the form, $\partial_{\mu}
 G^{\mu}$, in the standard model are studied. This  does not affect  the
 equation of motion in the bulk, but  makes  wave functions  modified,
 and causes the unusual transition characterized by the  time-independent 
probability.  
 In the  transition  probability  at a time-interval $T$ expressed 
generally in the form $P=T \Gamma_0 +P^{(d)}$, now with $
 P^{(d)}
 \neq 0 $. The diffractive term $P^{(d)}$ has the origin in the overlap of waves of 
the initial and final states, and reveals   the characteristics of waves. 
In
 particular, the processes of the neutrino-photon interaction ordinarily
   forbidden by Landau-Yang's theorem ($\Gamma_0=0$) manifests itself through
 the boundary interaction. 
 The new term leads  to  physical  processes  over a wide energy range 
to have finite probabilities.  New methods of detecting neutrinos using
 laser are  proposed that are based on this difractive term,
 which enhance the detectability of neutrinos by
 many orders of magnitude.

\end{abstract}
\newpage
\section{Matter wave  and $S[\text{T}]$} 
In modern science and technology, quantum mechanics plays
fundamental roles.   Despite the fact that stationary phenomena and 
method have been well developed, those of  non-stationary phenomena have
not.
In the former, the de Broglie wave length ${\hbar \over p}$ determines a
typical length  and  is of a smicroscopic size, and  scatterings or reactions
 in macroscopic scale    are considered 
independent, and successive reactions 
have been treated under 
the independent
scattering  hypothesis.
The probability  of the event that they  occur  
are computed by the incoherent sum of each value. In  the latter,
 time and space variables vary simultaneously, and a new scale, which
 can be much larger than the de Broglie wave length, emerges. 
They  appear  in overlapping regions of
the initial and final waves, and show  unique properties of intriguing
 quantum mechanical waves.
 
 A  transition rate   computed
 with a  method for stationary waves with   initial 
and final states defined at  the infinite time-interval
 $T= \infty$  is    independent 
of the details of the wave functions.  They hold characteristics 
of particles and preserve symmetry of the system.
 Transitions occurring at a finite $T$, however,
 reveal  characteristics  of waves, 
 the  dependence on the 
 boundary conditions \cite{ishikawa-tobita1,ishikawa-tobita2}
 \footnote{It was pointed out by Sakurai \cite{sakurai},
 Peierls \cite{peierls}, and 
Greiner \cite{greiner} that the probability at finite $T$ would be different
 from that at
 $T =\infty$. Hereafter we compute the difference for  the
 detected particle that has  the mean free path
 $l_{mfp}$ of  $l_{mfp}
 \gg c{T}$. },  and the probability,
\begin{eqnarray}
P=T \Gamma_0 +P^{(d)} \label{probability0},
\end{eqnarray}
where $P^{(d)}$ is the diffractive term which has often escaped
 attention by researchers.
 The rate $\Gamma_0$ 
is computed with  Fermi's golden rule \cite{Dirac,Schiff-golden,Goldberger,newton,taylor},
and preserves the internal and space-time symmetries, including the 
kinetic-energy conservation. $\Gamma_0$ holds  the characteristic properties
of particles
and the hypothesis 
of independent scatterings is valid.  For a particle  of
small mass, $m_s$, $\Gamma_0(p_i,m_s)$ behaves 
\begin{eqnarray}
\Gamma_0(p_i,m_s) \approx \Gamma_0(p_i,0),\label{scaling}
\end{eqnarray}
because  the characteristic  length, the de Broglie wave length,
 is determined with $p_i$. 
 The region where $P^{(d)}$ is ignorable is called the particle-zone.
  
 Overlapping   waves  in the initial and final states have the
finite interaction energy, and reveal  unique properties of waves
\cite{ishikawa-tobita1,ishikawa-tobita2}. 
Because  the interaction energy is part of the total energy, sharing 
with the kinetic energy,  the conservation law of the kinetic-energy is
    violated. Consequently, the state becomes non-uniform in time and
 the transition probability has a new component  $P^{(d)}$, showing 
  characteristics of waves. 
The term ${P^{(d)}
\over T}$ was shown to behave  with  a new scale of length, $({\hbar
\over m_s c}) \cdot({E_i \over m_s c^2})$. Accordingly the correction  is 
proportional to the ratio of two small quantities
\begin{eqnarray}
{P^{(d)} \over T} =f({1/T \over m_s^2 c^3/ {\hbar E_i} }),
\end{eqnarray}
and does not follow Eq. $(\ref{scaling})$.

$P^{(d)}$ reflects  the non-stationary waves and  is not computed 
with the stationary waves.    In the 
region where $P^{(d)}$ is important, the 
hypothesis of independent scattering is invalid, and interference
unique to waves manifests.  
This region is  called the wave-zone, and extends to a large area for 
light particles.
 $P^{(d)}$  has been ignored, but gives important contributions to the
 probability in
 various processes. 
Especially $P^{(d)}$ is  inevitable   for the process of $\Gamma_0=0$ and 
$ P^{(d)} \neq 0$, which  often appears. Furthermore,  $P^{(d)}$  can
be  enhanced  drastically, if the overlap of the waves is constructive
in wide area. 
This happens    for small $m_s$,  large $E_i$,  even for large $T$, and 
 reveals  macroscopic quantum phenomena.
 Processes of large $P^{(d)}$ involving  photon and 
neutrino  in the standard model are studied in 
the present paper.

An example of showing $\Gamma_0=0,\ P^{(d)} \neq 0$ is a system
of fields  described  
by a free part $L_0$ and  interaction part $L_{int}$ of 
total derivative,     
\begin{align}
L=L_0+L_{int},\ L_{int}={d \over dt} G,
\label{total-derivative}
\end{align}
where $G$ is a polynomial of fields $\phi_l(x)$.  $\phi_l(x)$  follows  
free equation,
\begin{eqnarray}
{\partial L_0 \over \partial \phi_l(x)}-{\partial \over \partial
 t}{\partial L_0 \over \partial {\partial \phi_l (x) \over \partial t}}=0.
\end{eqnarray}
   $L_{int}$  decouples from the equation and does not modify the 
equation of motion in classical  and quantum
mechanics. Nevertheless,
a  wave function $|\Psi(t)
 \rangle$   follows a Schr\"{o}dinger equation in the interaction picture,
\begin{eqnarray}
i\hbar{\partial \over \partial t}|\Psi\rangle_{int}= ({\partial \over
 \partial t} G_{int}(t))|\Psi \rangle_{int},
\label{Schroedinger}
\end{eqnarray} 
where the free part, $H_0$, and the interaction part, $H_{int}$, are derived
 from the previous Lagrangian, and $G_{int}$ stands for $G$ of the
 interaction picture.  
A solution at $t$, 
\begin{eqnarray}
|\Psi(t)\rangle_{int}=e^{G_{int}(t)- G_{int}(0) \over i\hbar}|\Psi(0)\rangle_{int},
\label{quasi-statioinary}
\end{eqnarray}
is expressed with   $G(t)$, and  the state  at $t > 0$ is   modified by the
interaction. The initial state
$|\Psi(0)\rangle_{int}$ prepared at $t=0$ is transformed to the other state
of t-independent weight. 
Hence,  
Eq. $(\ref{quasi-statioinary})$, is like stationary, and  
$\Gamma_0=0$ , and  $P^{(d)} \neq 0$.  
Physical observables are expressed by the probability of the events,
which are specified by the initial and final states. For those  at
finite $T$, normal S-matrix,
$S[\infty]$, which  satisfies the boundary condition at $T=\infty$
 instead of those at finite $T$,  is useless.  $S[T]$  that 
satisfies the boundary condition at $T$
\cite{ishikawa-tobita1,ishikawa-tobita2} is necessary and was 
constructed.  $S[T]$ is applied to 
 the system described by Eq. $({\ref{total-derivative}})$.
 
$S[T]$   is constructed with the M{\o}ller operators at a finite $T$,
$\Omega_{\pm}({T})$, as $S[T]=\Omega_{-}^{\dagger}(T)\Omega_{+}(T)$.
$\Omega_{\pm}({T})$  are expressed  by a free Hamiltonian  $H_0$ 
and  a  total Hamiltonian  $H$ by
  $\Omega_{\pm}(T)=\displaystyle\lim_{t \rightarrow \mp T/2}e^{iHt}e^{-iH_0t}$.
  From this expression,     
$S[T]$ is unitary and satisfies
\begin{align}
&[S[T],H_0] \neq 0,
\label{commutation-relation}
\end{align} 
hence   a  matrix
element of $S[T]$ between two eigenstates of $H_0$, $|\alpha \rangle$ 
and  $|\beta \rangle $ of eigenvalues $E_{\alpha}$ and $E_{\beta}$, 
is decomposed into  two components
\begin{eqnarray}
\langle \beta |S[{T} ]| \alpha \rangle=\langle \beta
 |S^{(n)}[{T} ]| \alpha \rangle+ \langle \beta |S^{(d)}[{T} ]|
 \alpha \rangle,
\label{matrix-element}
\end{eqnarray}
where $\langle \beta |S^{(n)}|\alpha \rangle$ and  $\langle \beta
|S^{(d)}|\alpha \rangle$ get contributions from  cases
$E_{\beta}=E_{\alpha}$  and  $E_{\beta} \neq E_{\alpha}$, and give
$\Gamma_0 \text T$ and $P^{(d)}$ respectively.
The deviation of the kinetic-energies, $E_{\beta} - E_{\alpha} $, in the
latter 
is due to the interaction energy of the
overlapping waves, which depends on the coordinate system. Therefore,
it is understood that   
  $H_{int}$ is not Lorentz invariant.    Thus  the kinetic-energy 
non-conserving term, which  was  
mentioned  by  Pierls and Landau \cite{peierls} so as to give 
negligibly small  correction, yields  $P^{(d)}$ \cite{greiner} 
\footnote{Unusual
enhancement observed in laser Compton experiment \cite{laser-compton}
may be connected.}. Because $H_0$ is an generator of the Poincare group, 
Eq.$(\ref{commutation-relation})$  shows that $S^{(d)}[\text T]$ and $P^{(d)}$
violate  the Poincare invariance.    
In the system described by Eq. $({\ref{total-derivative}}) $, the  first
term disappears but the second term does not,  $\Gamma_0=0$, and
$P^{(d)} \neq 0$.

$S[T]$
is  expressed with the boundary conditions for the
 scalar field $\phi(x)$ \cite{LSZ,Low},
\begin{eqnarray}
& &\lim_{t \rightarrow - T/2}\langle\alpha| \phi^f |\beta
 \rangle=  \langle \alpha| \phi_{in}^f |\beta \rangle,\\
& &\lim_{t \rightarrow + T/2}\langle \phi^f |0 \rangle=  \langle
 \alpha| \phi_{out}^f |\beta \rangle,
\end{eqnarray}
where $\phi_{in}(x)$ and $\phi_{out}(x)$ satisfy the free wave equation,
and $\phi^f,\ \phi_{in}^f,\ \text{and}\ \phi_{out}^f$ are the expansion
coefficient of  $\phi(x),\ \phi_{in}(x),\ \text{and}\ \phi_{out}(x)$, with the
normalized wave functions $f(x)$ of the form
\begin{eqnarray}
\phi^f(t)=i\int d^3 x f^{*}({\vec x},t) \overleftrightarrow{\partial_0}
 \phi({\vec x},t).
\end{eqnarray}
The function $f(x)$ indicates the wave function 
that the out-going wave interacts
in a successive reaction of the process. The out-going photon  studied in the following section interacts
with atom or nucleus and their wave functions are used for $f(x)$.
Consequently, $S^{(d)}[T]$ depends on  $f(x)$, and is appropriate
to write as  $S^{(d)}[T;f]$. 
   Accordingly the probability of the events   is 
expressed by this  normalized wave function, called wavepacket. 
Wavepackets that satisfy  free wave equations and are localized in space 
are important  for rigorously defining scattering amplitude  
\cite {LSZ,Low}.
 $S^{(d)}[T;f]$ expresses the wave nature due to  the states of 
continuous kinetic-energy. 
$S^{(d)}[T;f]$   does not preserve   
Poincar\'e invariance   defined by $L_0$. 
The state $|\beta \rangle $ of $E_{\beta}$ is orthogonal to 
$|\alpha \rangle $ of $E_{\alpha} \neq E_{\beta}$  and  
  $P^{(d)}$  approaches constant  at $T=\infty$.

The wavepackets \cite {LSZ,Low,Goldberger,newton,taylor}  
can be  replaced with plane waves for a practical computation for $S[\infty]$ 
\cite{Kayser,Giunti,Nussinov,Kiers,Stodolsky,Lipkin,Akhmedov,Asahara},
 but can not be done  so for $P^{(d)}$  
\cite{Ishikawa-Shimomura,Ishikawa-Tobita-ptp,Ishikawa-Tobita}.  
  $P^{(d)}$ is  derived from $S^{(d)}[T;f]$, and depends 
on $f(x)$. 

Photon is massless in vacuum and has a small effective mass  determined
by the plasma frequency in
matter, and neutrino  is nearly massless. Thus they   have the large
wave-zone of revealing wave 
phenomena caused by $P^{(d)}$. 
These small masses make  $P^{(d)}$
 appear in a macroscopic scale and  significantly affect physical
 reactions. In this small (or zero) mass region the effects of
 diffractive term $P^{(d)}$ are pronounced. This is our interest in the
 present paper. The
produced photon interacts with matter with
the electromagnetic interaction, which leads to macroscopic observables. 
The term $P^{(d)}$ of  the processes of $\Gamma_0=0$
such as  $2 \gamma$ decays of
$1^{+}$ meson and $\gamma$ and ${\nu}$ reactions 
are shown to be relevant to many physical processes, including  
possible experimental observation of relic neutrino. 
The enhancement of the
probability for light particles with intense photons
based on the normal component $\Gamma_0$ was proposed 
in Refs. \cite{homma-tajima1,homma-tajima2}, and the collective
interaction between electrons and neutrino derived from  the
normal component $\Gamma_0$ was considered in Ref. \cite{tajima-shibata}. 
Our theory is based on the probability  $P^{(d)}$, hence  differs from the
previous ones in many respects.

This  paper is organized  in the following manner. In Section 2, the 
couplings of two photon  with  $1^{+}$ state  through triangle diagram
is obtained. In section 3 and 4,  positronium and heavy quarkonium are
studied  and  their $P^{(d)}$ are computed.  
Based on these studies we go on to
investigate the interaction of photons and neutrinos. In section 5, 
neutrino--photon 
interaction of the order $\alpha G_F$, and  various implications to high energy
neutrino phenomena   are presented. In Section 6 we explore the
implication of the photon-neutrino coupling on experimental settings. Summary is given in section 7.

\section{Coupling of  $1^{+}$ meson  with two photons}
  The coupling of $\gamma \gamma$ with  axial vector states,  
 $1^{+}$ meson composed of $e^{-}e^{+}$, $Q \bar Q$,
and $\nu \bar \nu$   are studied  using  
an effective Lagrangian
  expressed by local fields.  
From  symmetry considerations, 
an effective  interaction of the $1^{+}$
 state  $\phi_1^{\mu}$ with two photons has the form,
\begin{eqnarray}
& &S_{int}=g \int d^4x \partial_{\mu} (\phi_1^{\mu}(x)\tilde F_{\alpha
 \beta}(x)F^{\alpha \beta}(x)),\label{effective-interaction}\\
& &\partial_{\mu} \phi_1^{\mu}(x)=0,\tilde F_{\alpha\beta}(x)=\epsilon_{\alpha\beta\gamma\nu}F^{\gamma\nu}(x), \nonumber
\end{eqnarray}
where $F_{\alpha\beta}$ is the electromagnetic field, and
the coupling strength  $g$ is computed later.
In a transition of plane waves in infinite-time
interval, the space-time boundary is at the infinity, and the transition 
amplitude  is 
computed with the plane waves, in  the form,
\begin{eqnarray}
M=(p_i-p_f)_{\mu}(2\pi)^4 \delta (p_i-p_f) \tilde M^\mu
\label{vanishing-gamma}
\end{eqnarray}
and vanishes, where $p_i$ and $p_f$ are  four-dimensional momenta of the
initial and final states and $\tilde M$ is the invariant amplitude. This
shows that  
the   amplitude proportional to
$\delta^{4}(p_i-p_f)$ and the transition rate $\Gamma_0$ vanish.
  The rate   of $1^{+} \rightarrow \gamma \gamma$ decay vanishes  
 in general  systems, 
because the state of two photons of momenta 
$(\vec{p}, - \vec{p})$ does not couple with a massive  $1^{+}$ particle.
 Hence  
\begin{eqnarray}
\Gamma_0^{1^{+}\rightarrow \gamma \gamma }=0,
\end{eqnarray}
which is  known as Landau-Yang's theorem \cite{landau,yang}.

The term $S_{int}$  is written as a surface term in four-dimensional
space-time,
\begin{eqnarray}
S_{int}=\int_{surface} d{ S_{\mu}} g (\phi_1^{\mu}(x)\tilde F_{\alpha
 \beta}(x)F^{\alpha \beta}(x)), 
\end{eqnarray}     
which is determined by  the wave functions of the initial and final states. 
Accordingly, the transition amplitude derived from this surface action is not 
proportional to $T$, but has a weaker $T$ dependence.
 Thus $P^{(d)}$ comes from the surface term, and  does 
not have   the delta function of  kinetic-energy conservation.
Kinetic energy of the final states deviates from that of the initial 
state due to the finite interaction energy between  them. The deviation
becomes larger and  $P^{(d)}$ is expected to increase with larger 
overlap. We find  $P^{(d)}$ in
the following.

\subsection{Triangle diagram}
\begin{figure}[t]
\begin{center}
\includegraphics[scale=.35,angle=0]{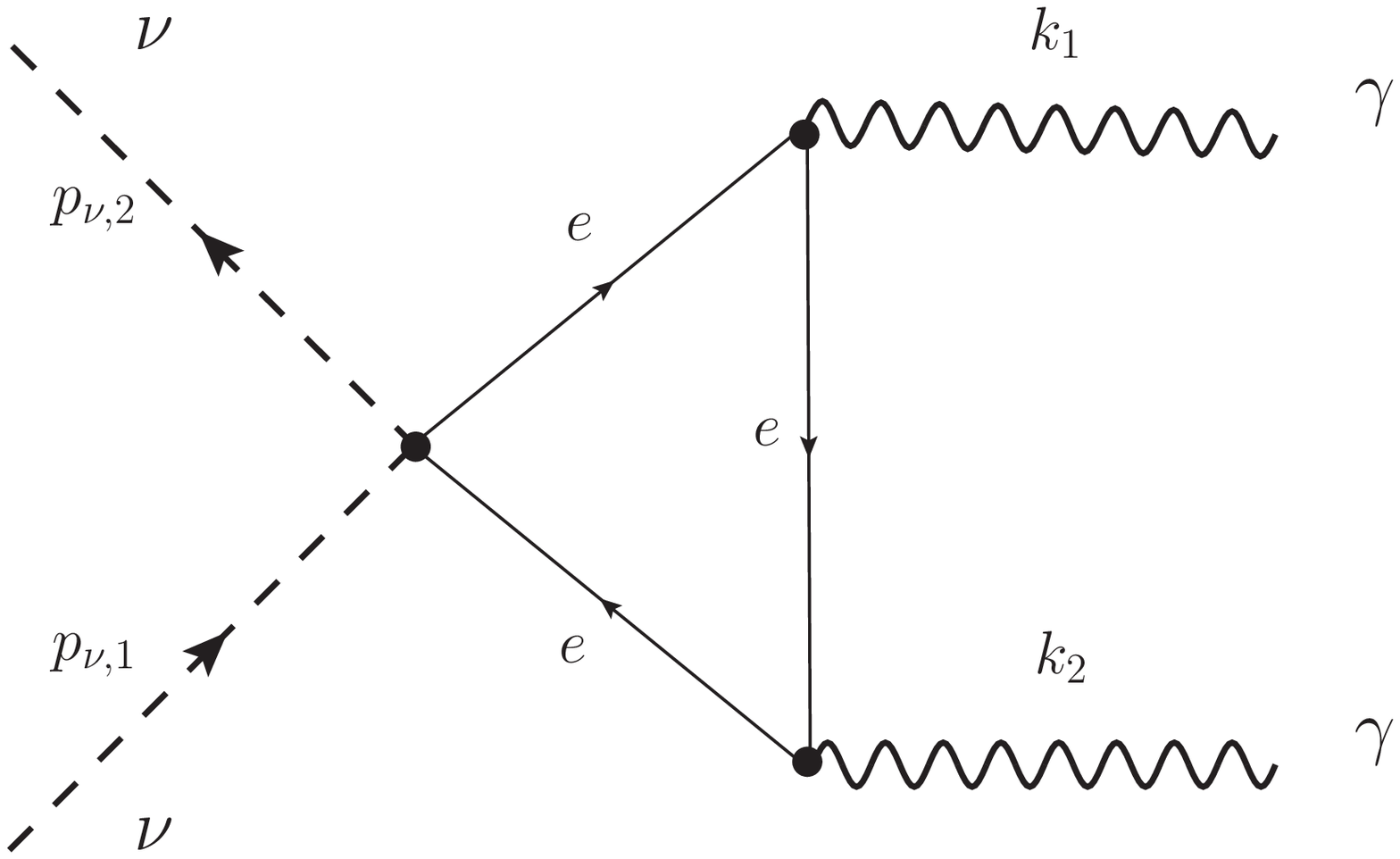}
\includegraphics[scale=.35,angle=0]{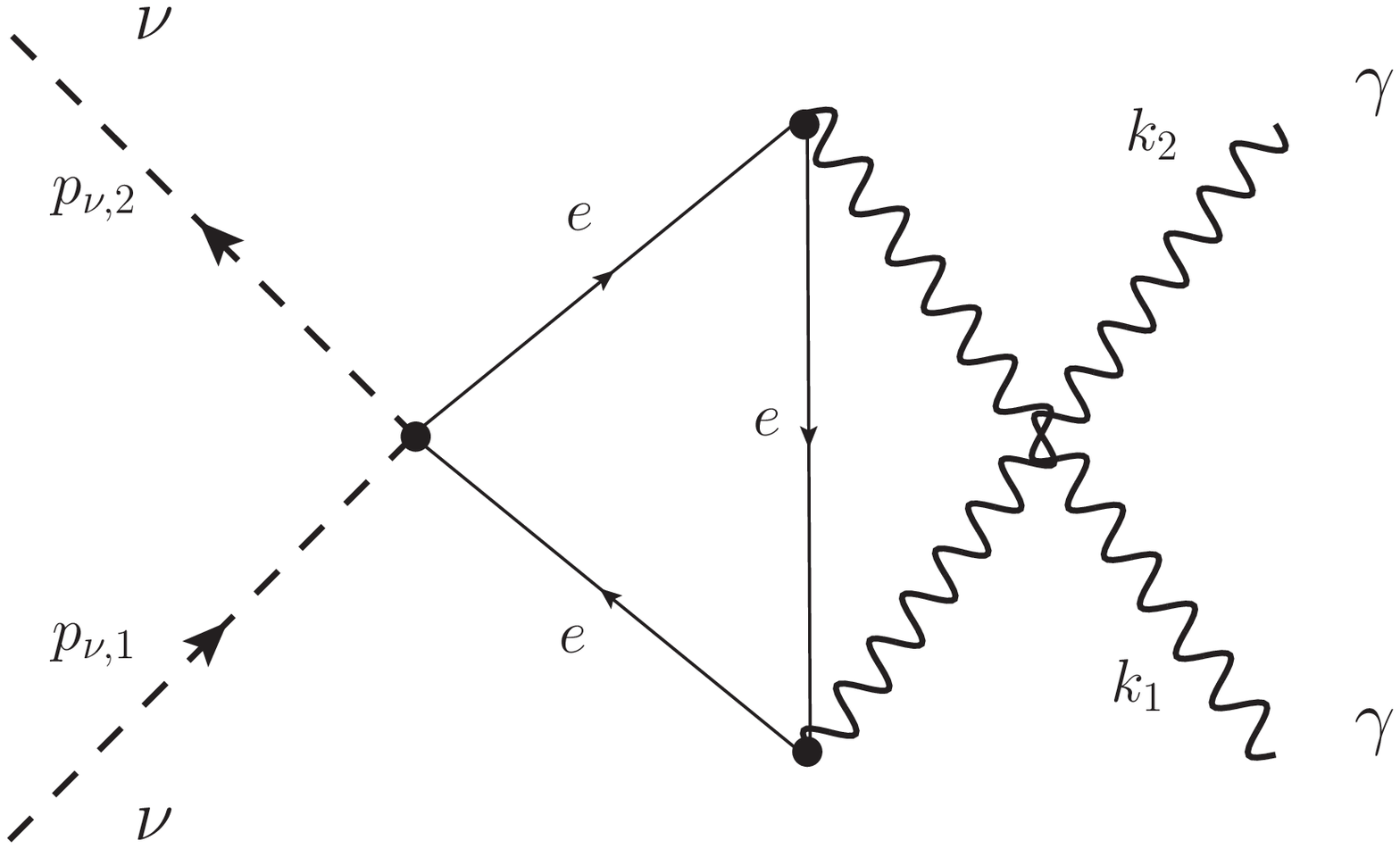}
\end{center}
\caption{Triangle diagrams of the electron loop which give contributions to 
$1^{+} (l \bar l) \rightarrow \gamma \gamma $, $\nu+\gamma \rightarrow \nu+\gamma$ and  $\nu+\bar \nu \rightarrow \gamma+\gamma$. 
  }
\label{tri-angle:fig}
\end{figure}

The interaction of the form Eq. $(\ref{effective-interaction})$ 
is generated  by  one loop effect in the standard model.  
The scalar and axial vector current 
\begin{eqnarray}
& &J(0)=\bar l(0) l(0),\\
& &J^{5,\mu}=\bar l(0) \gamma_5 \gamma^{\mu} l(0) 
\end{eqnarray}
in QED, we have 
\begin{eqnarray}
& &L=L_0+L_\text{int},\ L_0=\bar l(x)(\gamma\cdot p-m_l)l(x)-{1 \over
 4}F_{\mu\nu}(x)F^{\mu\nu}(x),\\
& &L_\text{int}=eJ_{\mu} A ^{\mu}(x),\ J_{\mu}(x)=\bar l(x) \gamma_{\mu} l(x), \nonumber
\end{eqnarray}
where $A^{\mu}(x)$ is the photon field and $l(x)$ is the electron field,
coupled through two photons in the bulk  through the triangle diagram 
Fig. \ref{tri-angle:fig}. The matrix elements  are
\begin{align}
\label{current-two-photon}
 &\Gamma_0=\langle 0|J(0) |k_1, k_2 \rangle =  {e^2 \over
 4\pi^2}\epsilon^{\mu}(k_1) \epsilon^{\nu}(k_2)
m [k_{2,\mu}k_{1\nu}-g_{\mu\nu}k_1 \cdot k_2] f_0,  \\
 &\label{axial-current-two-photon}
\Gamma_{5,\alpha}=\langle 0|J_{5,\alpha}(0) |k_1,k_2 \rangle =-i{e^2 \over
 4\pi^2} 2f_1 \epsilon^{\mu_1}(k_1) \epsilon^{\mu_2}(k_2)\nonumber \\
&\times [(k_{1,\mu_2}
\epsilon_{\mu_1\nu_1\nu_2\alpha}-k_{2,\mu_1}\epsilon_{\mu_2\nu_1\nu_2\alpha})k_1^{\nu_1}k_2^{\nu_2}
+(k_1\cdot
k_2)\epsilon_{\mu_1\mu_2\nu\alpha}(k_1-k_2)^{\nu}], 
\end{align} 
where $\epsilon_{\mu}(k)$ is the polarization vector for the photon. The
triangle diagram for the axial vector current
Eq. ($\ref{axial-current-two-photon}$) has been studied 
 in connection with axial anomaly and
 $\pi^0 \rightarrow \gamma \gamma$ decay
 \cite{fukuda-miyamoto,steinberger,rosenberg,adler,liu} and now is  
applied to $P^{(d)}$ for two photon  transitions  of the
 axial vector meson and  neutrino.  The triangle diagram Fig. 1 shows 
that the interaction occurs localy in space and time, but the transition
amplitude is the integral over the coordinates and receives the
large diffractive contribution if the  neutrino and
photon are spatially spread  waves.   In Fig. 1,
the in-coming and out-going waves are expressed by  lines, but they are
in fact the spread waves, which is obvious in the figures in Ref.
\cite{feynman} and in Fig. 2.

$\Gamma_{5,\alpha}$ is expressed also with $f_1$ in the form
\begin{eqnarray}
& &\Gamma_{5,\alpha}=\frac{\alpha_{em}}
 {2\pi}f_1(k_1+k_2)_{\alpha}(\tilde
 F_{1,\rho\lambda}F_2^{\rho\lambda}+\tilde
 F_{2,\rho\lambda} F_1^{\rho\lambda}),
\label{J_5-derivative}\\ 
&
 &F_{1,\rho\lambda}=k_{1,\rho}\epsilon_{1,\lambda}-k_{1,\lambda}\epsilon_{2,\rho},
 \tilde F_{\rho\lambda}={1 \over 2} \epsilon_{\rho\lambda\xi\eta}F^{\xi\eta}. \nonumber
\end{eqnarray} 

The coefficients $f_0$
 and $f_1$
 are given by the integral over the Feynman parameters,
\begin{eqnarray}
& &f_0=\int_0^1 dx \int_0^{1-x} dy{1 \over m_l^2-2xyk_1 \cdot
 k_2-i\epsilon},\\
& &f_1=\int_0^1 dx \int_0^{1-x} dy{xy \over m_l^2-2xyk_1 \cdot
 k_2-i\epsilon},
\end{eqnarray}
where 
\begin{eqnarray}
f_1
 &=&
-{1 \over 4k_1\cdot k_2}+{m_l^2 \over 4(k_1\cdot k_2)^2}I_1,\\
I_1&=&2k_1\cdot k_2\int_0^1dx \int_0^{1-x} dy{1 \over
 m_l^2-2xyk_1\cdot k_2-i\epsilon}     \nonumber\\
 &=&
\begin{cases}
2\text{Sin}^{-1}\left(\sqrt{k_1\cdot k_2\over 2m_l^2}\right) ,\ \text {if}~ k_1\cdot k_2
 < 2m_l^2,\\
 {\pi^2 \over 2}-{1 \over 2}\log^2{1+\sqrt{1-{2m_l^2 \over k_1\cdot k_2}}
 \over 1-\sqrt{1-{2m_l^2 \over k_1\cdot k_2}}}+i\pi \log{1+\sqrt{1-{2m_l^2
 \over k_1\cdot k_2}} \over  1-\sqrt{1-{2m_l^2 \over k_1\cdot k_2}}},\ \text {if}~k_1\cdot k_2
 \geq 2m_l^2.
\end{cases}
\end{eqnarray}
 $f_1$ in various kinematical regions is 
\begin{align}
f_1=
\begin{cases}
 {-1 \over 4k_1 \cdot k_2}+{m^2 \over 4(k_1 \cdot k_2)^2 } \pi(1-{2\delta E \over m_l});
\  k_1 \cdot k_2 =2m_l^2- m_l \delta
 E,\\
 {-1 \over 4k_1\cdot k_2};\ k_1 \cdot k_2 \gg m_l^2,\\ 
{1 \over 2m_l^2};\ k_1 \cdot k_2 \ll  m_l^2.
\end{cases}
\end{align}
\begin{figure}[t]
\begin{center}
\includegraphics[scale=.6,angle=0]{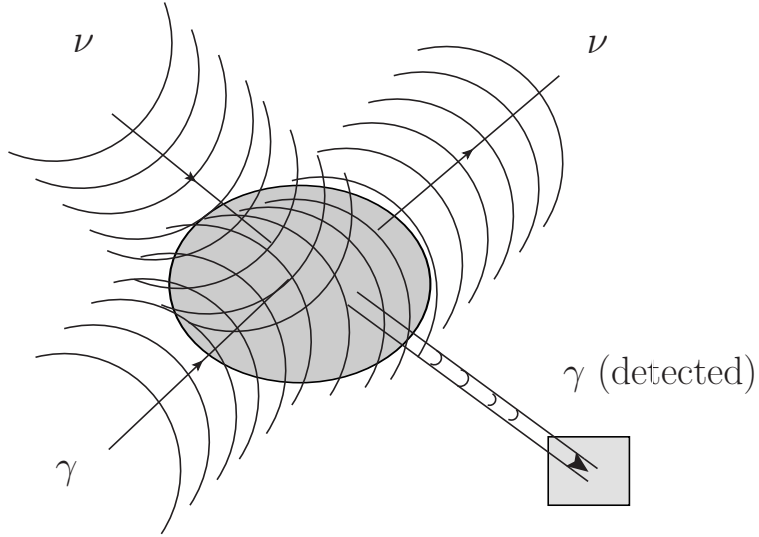}
\end{center}
\caption{Diagram of the neutrino photon scattering, $\nu+\gamma
 \rightarrow \nu+\gamma$ of spatially spread waves.  }
\label{neutrino-gamma:fig}
\end{figure}

\section{ Positronium   }

The bound states of a positronium  with the orbital angular momentum 
 $L=1,S=1$    have  total angular momentum $J=2,1,0$.  These states 
at rest of ${\vec P}=0$ are expressed with  
non-relativistic wave functions and   creation operators of $l^{+}$
and $l^{-}$ of momentum ${\vec p}$, spin $\pm 1/2$, and   P-wave  wave 
functions  $p_i F(p)$ as are given in Appendix A, where 
\begin{eqnarray}
| 1/2,1/2; -1 \rangle  =\int d{\vec p}\, b_{+1/2}^{\dagger}({\vec
 p}\,)d_{+1/2}^{\dagger}(-{\vec p}\,)(p_x-ip_y)F(p) |0 \rangle .
\end{eqnarray}
 Others are defined in the same manner.  

These bound states couple with the lepton pairs with effective local 
interactions which are expressed as
\begin{eqnarray}
L_{int}= g_0  \phi_0(x) \bar l(x) l(x)+ g_1 \phi_1^{\mu}(x) \bar l(x) \gamma^5
 \gamma_{\mu}l(x)+g_2 \phi_2^{\mu\nu}(x) \bar l(x)
 \gamma_{\mu}\partial_{\nu}l(x),
\end{eqnarray}
where the coupling strengths are computed from
 \begin{eqnarray}
& & g_0 =\langle 0|\bar l(0) l(0)|\phi_{0};{\vec p}=0  \rangle, \\
& &  g_1  \epsilon^{\mu}({\vec p}=0)= \langle 0|\bar l(0)\gamma^{5}
 \gamma^{\mu}l(0)|\phi_{1}; {\vec p}=0\rangle,\\
& & g_2\epsilon^{\mu\nu}({\vec p}=0)=\langle 0|\bar l(0)
 \gamma^{\mu}{\partial}^{\nu} l(0)|\phi_2 ;\vec p=0 \rangle,
\end{eqnarray}
where  the $\epsilon^{\mu},\epsilon^{\mu\nu}$ are polarization vector or
tensor for the massive vector and tensor mesons. We have 
\begin{eqnarray}
& &g_0=gN_0,g_1=gN_1,g_2=gN_2,  \label{couplings}\\
& &g=-2\pi \int dp\, p^4 F(p){2 \over |E|+m}, \nonumber\\
& &N_0=1,\ N_1=0.862,\ N_2=1.33. \nonumber 
\end{eqnarray}
The decay rates for $0^{+}$ and $2^{+}$ were  studied in
\cite{aleskseev, tumanov}, so that here we
concentrate on $1^{+}$ and $0^{+}$ as a reference for $1^{+}$.  

Fields $\phi_0(x),\phi_1(x), \text{ and } \phi_2(x)$ couple with 
two photons through the triangle diagrams. Their interactions   with 
two photons  are summarized in the following effective Lagrangian,
\begin{eqnarray}
& &L_{int}=g_0 {\alpha \over \pi} f_0 \phi_0 F_{\mu\nu} F^{\mu \nu} +g_1
 {\alpha \over \pi} f_1 \partial_{\mu}
 (\phi_1^{\mu}\epsilon_{\nu\rho\sigma\tau}F^{\nu\rho}F^{\sigma\tau})+g_2{\alpha
 \over \pi} 
T^{\mu\nu} F_{\mu\rho}F^{\rho\nu}.
\label{effective-two-photon} \nonumber \\
& &
\end{eqnarray}  

 \subsection{Axial vector positronium}   
Here we study  the two photon decay of axial vector positronium,
which is governed  by 
the second term of the right-hand side of
Eq. $(\ref{effective-two-photon})$.  
The matrix element of the axial current between the vacuum and two
photon state was  computed by Refs. 
\cite{fukuda-miyamoto,steinberger,rosenberg,adler,liu}. Because $\Gamma_0$
of two photon decay of axial vector meson vanishes due to Landau-Yang's 
theorem, but $P^{(d)}$ does not, we give the detailed derivation of $P^{(d)}$.

From the effective  interaction, Eq. $(\ref{effective-two-photon})$,   
the probability amplitude of the event that one of the photons of ${\vec
 k}_{\gamma}$  from
the decay of $c_{\mu}$ of $\vec p_c$ is detected at $({\vec X}_{\gamma},T_{\gamma})$  is   
\begin{align}
\mathcal M=-g_1{\alpha \over \pi} \int d^4x {\partial \over \partial
 x_{\mu}} [ f_1\langle 0|c_{\mu}(x)|{\vec p}_c \rangle \langle ({\vec
 k}_{\gamma},{\vec
 X}_{\gamma},T_{\gamma}),k_1|\epsilon_{\nu\rho\sigma\tau}
 F^{\nu\rho}F^{\sigma\tau}|0 \rangle], \label{amplitude} 
\end{align}
where 
\begin{align}
&\langle {\vec k}_{\gamma},{\vec
 X}_{\gamma},T_{\gamma}|A^{\mu}(x)|0\rangle  \nonumber\\
&=N_{\gamma} \int
 d{\vec k}_2 \rho_{\gamma}({\vec k}_2)  e^{-{\sigma_{\gamma}\over 2}({\vec k}_2-{\vec
 k}_{\gamma})^2+i(E({\vec k}_2)(t-T_{\gamma})-{\vec k}_2({\vec
 x}-{\vec X}_{\gamma}))} \epsilon^{\mu}({\vec k}_2), \label{couplings1}\\
&\langle {\vec k}_{1}|A^{\mu}(x)|0\rangle = \rho_{\gamma}({\vec k}_1) \epsilon^{\mu}({\vec k}_2)e^{{+i(E({\vec k}_1)t-{\vec k}_1{\vec
 x})}}, \nonumber \\
&\langle 0|c_{\mu}(x)|{\vec p}_c \rangle=( 2\pi)^{3/2} \rho_c({\vec p}_c)\epsilon^{\mu}({\vec
 p}_c)e^{{-i(E({\vec k}_c)t-{\vec k}_c{\vec
 x})}}. \nonumber
\end{align}
 The  initial state is normalized, and 
  the coupling in Eq. $(\ref{couplings})$   has 
$(2\pi)^{\frac{3}{2}}$ for the initial state, and    
\begin{eqnarray}
 & &k_{\mu} \epsilon^{\mu}(k)=0,\ (p_c)_{\mu} \epsilon^{\mu}(p_c)=0, \\
& &\rho({\vec k})=\left({1 \over 2E(k)(2\pi)^3}\right)^{\frac{1}{2}}. \nonumber
\end{eqnarray}  
The  state $  \langle {\vec k}_{\gamma},{\vec
 X}_{\gamma},T_{\gamma}| $ is normalized and, 
\begin{eqnarray}
N_{\gamma}=\left({\sigma_{\gamma} \over \pi}\right)^{\frac{3}{4}}.
\end{eqnarray}

Integration  over ${\vec k}_2$ is made prior to the integration
over $x$, in order for $\mathcal M$ to satisfy the boundary condition of
$S[T]$, and we have,
\begin{eqnarray}
& &\langle {\vec k}_{\gamma},{\vec
 X}_{\gamma},T_{\gamma}|A^{\mu}(x)|0\rangle
=\theta(\lambda)\left({(2 \pi)^3 \over
 \sigma_{\gamma} \sigma_T^2}\right)^{\frac{1}{2}} 
\rho({\vec k}_{\gamma}+\delta {\vec k} )\epsilon_{\mu}({\vec
k}_{\gamma}+\delta {\vec k}) \nonumber\\
& & \times e^{i(E({\vec k}_{\gamma})(t-T_{\gamma})-{\vec k}_{\gamma}({\vec
 x}-{\vec X}_{\gamma})) -\chi({ x_{\mu}})},\label{stationary-equation}
\end{eqnarray}
where 
\begin{eqnarray}
& &\lambda=(t-T_{\gamma})^2-({\vec x}-{\vec X}_{\gamma})^2,\\
& &\chi(x)={1 \over 2\sigma}_{\gamma}(({\vec x}-{\vec
 X}_{\gamma})_l-{ v}_{\gamma}(x_0-T_{\gamma}))^2+{1 \over 2\sigma_{T}}({\vec x}-{\vec
 X}_{\gamma})_T^2,  \\
& &(\delta {\vec   k}(x))_i=-{i \over \sigma_{\gamma}^i} \delta {\vec  x},
\delta {\vec x}=({\vec x}-{\vec
 X}_{\gamma}-{\vec v}_{\gamma}(t-T_{\gamma})),\nonumber \\
& &\sigma_{\gamma}^l=\sigma_{\gamma};\ i=\text{longitudinal},\ 
\sigma_{\gamma}^T=\sigma_{T};\ i=\text{transverse}.\nonumber
\end{eqnarray}
The wavepacket expands  in the transverse direction, and $\sigma_T$ in large 
$x_0-T_{\gamma}$ is given by
\begin{eqnarray}
\sigma_T=\sigma_{\gamma}- {i  \over E}(x_0-T_{\gamma}).
\end{eqnarray}
We later  use 
\begin{eqnarray} 
& &   \sum_{spin} \epsilon_{\mu}({\vec k}_{\gamma}+\delta {\vec k} )
 \epsilon_{\nu}({\vec k}_{\gamma}+\delta {\vec k} )=-g_{\mu \nu},\\
& &\delta    k^0(x)={i \over \sigma_{\gamma}} \delta  x^{0},\ \delta x^{0}=
 {\vec v}({\vec x}-{\vec
 X}_{\gamma}-{\vec v}_{\gamma}(t-T_{\gamma})). \nonumber
\end{eqnarray}

The stationary phase  for large $x_0-T_{\gamma}$ exists in the time-like
region $\lambda \geq 0$ \cite{Ishikawa-Shimomura}, and the function and
its derivative is proportional to $\theta(\lambda)$.  Thus the
integration  in Eq. $(\ref{amplitude})$ is made over 
the region
 $ \lambda \geq 0$, which has the boundary, $\lambda=0$.  
Consequently the transition amplitude does not
vanish if the integrand  at the boundary is finite. It is shown that this
is the case in fact. 
${\sigma_{\gamma}}$ is the size
of the nucleus or atomic  wave function that the photon interacts with 
and is  estimated later.
For the sake of simplicity, we use the Gaussian form for the main part 
in this paper.

Substituting Eq. $(\ref{axial-current-two-photon})$, we have  the amplitude
\begin{eqnarray}
& &\mathcal M=-ig_1{\alpha \over \pi} N \int_{\lambda \geq 0} d^4x 
{\partial \over \partial x_{\mu}}[e^{i(p_c-k_1)x} 
      \tilde {\mathcal M}^\mu ], \label{amplitude-axial-vector}\\
& &\tilde {\mathcal M}^\mu= f_1\{k_1\cdot
( k_{\gamma}+\delta k(x))\}e^{i(E(k_{\gamma})(t-T_{\gamma})-{\vec k}_{\gamma}({\vec
 x}-{\vec X}_{\gamma}))-\chi(x)}  T^{\mu},
 \nonumber \\
& &T^{\mu}=  \epsilon^{\mu}(p_c) \epsilon^{\nu}(k_1)^{*} \epsilon_{\rho\lambda\xi\nu}k_1^{\xi}({ k}_{\gamma}+\delta
 k_{\gamma}(x))^{\rho} \epsilon^{\lambda}(k_{\gamma}+\delta
 k_{\gamma}(x)), \nonumber \\
& &N=(2\pi)^{3/2}\rho_c({\vec p}_c)\rho_{\gamma}({\vec k}_{\gamma}) \rho({\vec k}_1) N_{\gamma} 
\left({(2\pi)^3 \over
   \sigma_{\gamma} \sigma_T^2}\right)^{\frac{1}{2}}, 
\end{eqnarray}
which   depends on the momenta  and 
 coordinates $({T}_{\gamma},\vec{{X}}_{\gamma})$ of the final state
and ${T}_{m}$.  
Although  $\mathcal M $ is written as the integral over the surface,  
$\lambda = 0$, the expression
 Eq. $(\ref{amplitude-axial-vector})$ is useful and applied for computing 
the  probability per  particle 
\begin{eqnarray}
P={1 \over V} \int {d\vec{{X}}_{\gamma} \over (2\pi)^3} d{\vec k}_{\gamma}  d{\vec
 k}_1 \sum_{s_1,s_2}|\mathcal M|^2,
\label{total-probability-gamma} 
\end{eqnarray}
 where $V$ is a normalization volume for the initial state,  and  the
 momentum of the non-observed final state  is integrated over the
whole positive energy region and the  position  of the
observed particle are integrated inside the detector. 

  Following 
the method of the previous works
\cite{ishikawa-tobita1,ishikawa-tobita2} 
and  Appendix B, we write the probability with a correlation function.
In the integral   
\begin{align}
\int d{\vec k}_1 |\mathcal M|^2=& 
\left({g\alpha N \over \pi \rho(k_1)}\right)^2 
\int d{\vec k}_1\rho^2({\vec k}_1)\int_{\lambda_1 \geq 0} d^4x_1 d^4x_2 {\partial^2 \over
 \partial_1^{\mu_1} \partial_2^{\mu_2}}F\label{probability-integral},\\
F=&f_1\{k_1(k_{\gamma}+\delta
 k_{\gamma}(x_1))f_1^{*}(k_1(k_{\gamma}+\delta k_{\gamma}(x_2)) )\}
 T^{\mu_1}(x_1) T^{\mu_2,*}(x_2) \nonumber \\
&\times e^{-i(p_c-k_1-k_{\gamma})(x_1-x_2)-\chi(x_1)-\chi^{*}(x_2)}, \nonumber
\end{align}
we have
\begin{align}
&\sum_{spin} T^{\mu_1}(x_1) T^{\mu_2,*}(x_2) =\nonumber\\
&2 (-g^{\mu_1\mu_2}+p_c^{\mu_1}p_c^{\mu_2}/{M^2})
(k_1(k_{\gamma}+\delta k_{\gamma}(x_2)^{*}))(k_1(k_{\gamma}+\delta
 k(x_1))). \label{squareM}
\end{align}
With    variables 
\begin{eqnarray}
& &x_{+}^{\mu}={x_1^{\mu}+x_2^{\mu} -2 X^{\mu} \over 2},\ X_0=T_{\gamma}, \\
& & \delta x^{\mu}=x^\mu_1-x^\mu_2,
\end{eqnarray} 
the
integral is written as  
\begin{align}
\int_{\lambda_i \geq 0} d^4x_1 d^4x_2 {\partial^2 \over \partial x_1^{\mu} \partial
 x_{2,{\mu}}}F 
&=\int_{\lambda_{+} \geq 0} d^4x_{+} d^4 \delta x \left({1 \over 4}{\partial^2
 \over {\partial  x_{+,\mu}}^2 }-{\partial^2
 \over {\partial  {\delta x}_{\mu} }^2}\right)F \nonumber\\
&=\int d^4 \delta x \int_{\lambda_{+}=0} {1 \over 4}d^3S_{+}^{\mu}{\partial
 \over {\partial  x_{+,\mu}} } F, \label{integral}\\
\end{align} 
where $\lambda_{+}=x_{+}^2$ and $x_{+}$ 
is  integrated over the region
$\lambda_{+} \geq 0$, and the integral is computed with the value at
the boundary, $\lambda_{+}=0$.  $\delta x$  are integrated over the whole 
region, and the second term in the second line  vanishes.

Using  the  formulas \cite{ishikawa-tobita1,ishikawa-tobita2},  
\begin{eqnarray}
& &{1 \over (2\pi)^3} \int {d{\vec k}_1 \over
 2E(k_1)}e^{-i(p_c-k_1)\cdot \delta x} =-
i{1 \over 4\pi}\delta(\lambda_{-})\epsilon(\delta t)
\theta(\text{phase-space})+\text{regular},  \nonumber \\
& &
 \theta(\text{phase-space})=\theta(M^2-2p_cp_{\gamma}), 
\label{phase-space}\\
& &\int d \delta {\vec x}e^{ip_{\gamma}{\delta {\vec x}}-{1 \over
 4\sigma}(\delta {\vec x}-{\vec v}\delta t)^2}{1 \over
 4\pi}\delta(\lambda)=
{\sigma_T \over 2 } {e^{i\bar \phi_c(\delta t)} \over \delta t},
\label{light-cone-integration}\\
& &\ \bar \phi_c(\delta t)= \omega_{\gamma} \delta t,\omega_{\gamma} ={m_{\gamma}^2 \over
  2E_{\gamma}}, \nonumber
\end{eqnarray} 
and the integrals given in Appendix E,  
we compute the probability. 
It is worthwhile to note that  the right-cone singularity exists
 in the kinematical region  $\theta(\text{phase-space})$
and the probability becomes finite in this region 
\cite{ishikawa-tobita1,ishikawa-tobita2}. 
The natural unit, $c=\hbar=1$, is taken in the majority of places, 
but $c$ and $\hbar$ are written explicitly 
when it is necessary, and MKSA unit is taken in later parts. 
After  tedious calculations, we have the probability
 \begin{align}
P
&= {1 \over 3} \int {d\vec{\text{X}}_{\gamma} \over V}{d{\vec
  p}_{\gamma} \over (2\pi)^3 E_\gamma}  N_2 {1 \over 4\pi} \left(i\sum_i I_i\right) 
\Delta_{1{+},\gamma} \theta(M^2-2p_c\cdot p_{\gamma}),
 \label{probability-correlation}   
\end{align}
where $I_i$ is given in Appendix E, and 
\begin{align}
&N_2=(g{2 \over \pi}\alpha )^2 
{1 \over 2E_c} ({\pi \over \sigma_{\gamma}})^{3/2},\\
&\Delta_{1{+},\gamma} 
= |f_1((p_c-k_{\gamma})\cdot k_{\gamma})|^2 2((p_c-k_{\gamma} ) \cdot
k_{\gamma})^2 ={(\pi-2)^2 \over 32};\ (\delta E \rightarrow 0),\label{1^{+}-photoncorrelation}
\end{align}
and 
\begin{eqnarray}
& &\left({\pi^3 \sigma_{\gamma} \over  \sigma_T^2(1)
 {\sigma_T^2(2)^*}}\right)^{\frac{1}{2}}=
\left({\pi^3 \over
 \sigma_{\gamma}^3}\right)^{\frac{1}{2}}\rho_s(x_1^0-x_2^0)\label{spreading},\\
& &\rho_s(x_1^0-x_2^0)=1+i{x_1^0-x_2^0 \over
 E_{\gamma}\sigma_{\gamma}}\nonumber
\end{eqnarray}
were substituted. 

Integrating over the gamma's coordinate $\vec{{X}}_{\gamma}$,
we obtain  the total volume, which   is canceled by the factor 
$V^{-1}$ from  the normalization of the initial state.
The total probability, for the high-energy gamma rays,  is then expressed as
\begin{align}
\label{probability-3}
P= {1 \over 3}  \pi^2\left\{\sigma_l^2 \text {log}(\omega T)+{\sigma_l \over m_{\gamma}^2}\right\}
  \int \frac{d^3 p_{\gamma}}{(2\pi)^3 E_\gamma} {1
 \over 2} N_2  
\theta(M^2-2p_c \cdot p_{\gamma} )\Delta_{1{+},\gamma} ,
\end{align}
where  $\text{L} = c{T}$ is the
length of the decay region.  The
kinematical region of the final states is expressed by the step function
\cite{ishikawa-tobita2}, which is different from the on-mass shell
condition. $\Gamma_0$ vanishes and $P$ is composed of   
$\log T$  and constant terms.  The constant  is inversely proportional 
to $m_{\gamma}^2$, and becomes large for the small $m_{\gamma}$.

\subsection{  Transition probability }

A photon $\gamma$ is massless in vacuum and has an effective mass in matter,
 which  is 
given by plasma frequency in matter \cite{tajima-dawson} as  
\begin{eqnarray}
m_{eff}=\hbar \sqrt {n_e e^2 \over m_{e} \epsilon_0}=\hbar\sqrt{4\pi
 \alpha n_e \hbar c \over m_e},
\end{eqnarray}
where  $n_e$ is the  density of electron, and for the value,   
\begin{eqnarray}
n_e^{0}={1 \over (10^{-10})^3} (\text{m})^{-3}=10^{30} (\text{m})^{-3}.
\end{eqnarray}
 Substituting
\begin{eqnarray}
 \alpha={1 \over 137},\ m_e=0.5\,\text{MeV}/c^2,
\end{eqnarray}
we have,
\begin{eqnarray}
m_{eff}c^2=30 \sqrt{n_e \over n_e^{(0)}}\, \text{eV}.
\end{eqnarray}
In the air, $ n_e= 3\times 10^{25} /\text{m}^3$, the mass agrees with
\begin{eqnarray}
 m_{eff}c^2=4\times 10^{-3}\, \text{eV},
\end{eqnarray} 
which is comparable to the neutrino mass.
At a macroscopic $T$, 
\begin{eqnarray}
\omega T=T/T_0,\ T_0^{-1}={m_{eff}^2 c^4 \over \hbar E}  =900{n_e \over
 n_e^{(0)}}{1 \over 6 \times 10^{-16}\times 10^6}\text{sec}^{-1} ,
\end{eqnarray}
$\omega T$ is much larger than 1.

We have the probability
at the system of $p_c=(E_c,0,0,p_c)$, 
\begin{eqnarray}
& &{d P \over dp_{\gamma}}=  C\left((1-{E_c \over p_c}) p_{\gamma} +{M^2 \over 2p_c}\right);\ 
{M^2 \over 2(E_c+p_c)}
 \leq p_{\gamma} \leq {M^2 \over
 2(E_c-p_c)}, \\
& & C={\pi^2 \over 3}\left(3 \sigma_\gamma^2 \log \frac{T}{T_0} +{1 \over 4}{\sigma_{\gamma} \over
 m_{\gamma}^2}\right){N_2 \over 2} \Delta_{1+.\gamma}, \nonumber 
\end{eqnarray}
and    the total probability 
\begin{eqnarray}
P_{total}={(\pi-2)^2 \over 1536}{1  \over \sqrt \pi}
\left(3 \sqrt
 \sigma_{\gamma} \log\frac{T}{T_0}+{1 \over 4}{1 \over m_{\gamma}^2 \sqrt \sigma_{\gamma}}\right)
(g \alpha )^2 
{E_c+2p_c  \over E_c(E_c +p_c)}.\label{positoronium-two-photon}
\end{eqnarray}
   
\section{Heavy quarkonium }
The decay of axial vector mesons composed  of heavy quarks exhibits the same 
phenomena. The heavy quark mesons  composed of charm and bottom quarks 
are  observed, and show rich decay properties   in the two photon decay,
radiative transitions, and two gluon decays.    Because quarks 
interact via  electromagnetic and strong interactions,
  non-perturbative effects are not
negligible, but the symmetry consideration is valid. Moreover, the 
non-relativistic representations are  good for these
bound states, because quark masses are much greater  than the confinement
scale.  Furthermore they have small spatial sizes.  Accordingly  we
represent them with local fields and find their interactions using 
the coupling strengths of Eq. $(\ref{couplings})$ and the
values of the triangle diagrams, Eqs. $(\ref{current-two-photon})$, 
and (\ref{axial-current-two-photon}).
\subsection{$ q \bar{q} \rightarrow \gamma + \gamma$} 
Up and down quarks have charge $2e/3 $ or $e/3$, and color triplet.
Hence the probabilities of  two photon decays  are obtained by the expression 
Eq. $(\ref{positoronium-two-photon})$ with charges of quarks ${2e/3},\ {e/ 3}$,  and color
factor. It is highly desired to obtain  the experimental value for $1^{+}$. 
 
\subsection{$q \bar{q} \rightarrow gluon+ gluon$}

A meson composed of heavy quarks decays to light hadrons through
gluons.  Color singlet two gluon states are equivalent to two
photon states. Accordingly  the two gluon decays  are  calculated 
in the equivalent way to that of two photon decays as far as 
the perturbative calculations are concerned. The transition rates  for 
$0^{+}$ and $2^{+}$ may be calculated in this manner.  The total rates for 
$L=0$ charmonium,  $J/\Psi$ and $\eta_c$   agree with the values
obtained by the perturbative calculations. $J/\Psi$ is $C=-1$  and decays 
to three gluons and the latter is $C=+1$ and decays to two gluons.
The former rate is of $\alpha_s^3 $ and  the latter rate is of $\alpha_s^2$, where
$\alpha_s $ is the coupling strength of gluon. Their
widths are  $\Gamma= 93$ keV or $\Gamma= 26.7$ MeV, and  are consistent with
small coupling strength $\alpha_s\approx 0.2$.  Now  $L=1$ states have $C=+1$.
Hence  a meson of $J=2$ and that of $J=0$ decay to two gluons, whereas 
the decay rate of a $J=1$ meson
 vanishes  by Landau-Yang's theorem.  

A $c\bar{c}$ meson of the quantum number  $1^{+}$  is slightly different from that of
positronium, because a gluon  hadronizes by a non-perturbative effect,
which is peculiar to the gluon.  The hadronization length is not rigorously 
known, but it would be reasonable to assume that the length is on the 
order of the size of pion.  The time interval $T$ is then a microscopic
value.  $P^{(d)}$ for this time $T$ is estimated in the following.

A gluon  also hadronizes in the interval of the lightest hadron  size,    
\begin{eqnarray}
T={R_{hadron} \over c}={c \hbar \over m_{\pi}}.
\end{eqnarray}
The gluon plasma frequency is estimated from quark density, 
\begin{eqnarray}
m_{eff}=\sqrt {n_q e_{strong}^2 \over m_q \epsilon_0},
\end{eqnarray}
and
\begin{eqnarray}
n_q={1 \over (\text {fm})^3},\ \alpha_s=0.2,\ m_q c^2=2 \text{ MeV},
\end{eqnarray}
 Then we have
\begin{align}
\omega T&={ m_{eff}^2 c^4 T \over E}=\frac{c^4 \alpha_s^2}{(\text{fm})^3 m_q m_{\pi} E} ={(c\hbar)^3 \over m_q
 m_{\pi} E (\text{fm})^3}\alpha_s^2 \nonumber\\
&={(197\text{MeV})^3 \over 2\times 130\times
 1500(\text{MeV})^3}\times 4\times 10^{-2} =0.9.
\end{align}
At small  $\omega T$, $\tilde g(\omega T)$ varies with $T$ as is shown in
Fig. 2 of Ref. \cite{ishikawa-tobita2} and    
\begin{eqnarray}
\tilde g(\omega T)|_{\omega T=1} = 2.5.
\end{eqnarray}
 
Higher order corrections also modify the rates for $q\bar{q}$. The
values to light hadrons  are
estimated by \cite{barbieri, kwong,li,bodwin,huang}. The total values for light hadrons are expressed
with a singlet and octet components $H_1$ and $H_8$ as  
\begin{eqnarray}
& &\Gamma(\chi_0 \rightarrow \text {light~ hadrons})=6.6\alpha_s^2 H_1+3.96H_8\alpha_s^2,\\
& &\Gamma(\chi_1 \rightarrow  \text { light ~hadrons})=0\times \alpha_s^2 H_1+3.96H_8\alpha_s^2,\\
& &\Gamma(\chi_2 \rightarrow  \text {light ~hadrons})=0.682\alpha_s^2 H_1+3.96H_8\alpha_s^2.
\end{eqnarray}
Using values for $\chi_0$ and $\chi_2$ from \cite{particle-data}
\begin{eqnarray}
& &\Gamma(\chi_0 \rightarrow \text{light hadrons})=  1.8 \ \text{MeV},\\
& &\Gamma(\chi_2 \rightarrow \text{light hadrons})= 0.278\ \text{MeV},
\end{eqnarray}
 we have the rate   for $\chi_1$ from $H_8$,
\begin{eqnarray}
\Gamma^{(8)}(\chi_1 \rightarrow  \text {light~ hadrons})=  0.056 \text { MeV}.
\end{eqnarray}
The experimental value for $\chi_1$ is 
\begin{eqnarray}
& &\Gamma(\chi_1 \rightarrow  \text {light~hadrons })= 0.086  \text{ MeV (world ~average)},
 \label{world-average-chi_1}\\
& &\Gamma(\chi_1 \rightarrow  \text {light~hadrons})= 0.139  \text{ MeV (BESS II)} .
\label{Bess-chi_1}
\end{eqnarray}
The large discrepancy among  experiments  for $\Gamma_{\chi_1 \to \text{light hadrons}}$ may suggest that
 the value depends on the experimental situation, which may be a feature of 
$P^{(d)}$.  We have
\begin{eqnarray}
\Delta \Gamma=\Gamma(\chi_1 \rightarrow  \text {light~hadrons })-\Gamma^{(8)}(\chi_1
 \rightarrow  \text {light~ hadrons}),
\end{eqnarray}
\begin{eqnarray}
& &\Delta \Gamma( \text {world ~average})=  0.030 \text{ MeV },\\
& &\Delta \Gamma( \text {BESSII })=  0.083  \text{ MeV}, \nonumber
\end{eqnarray}
which are attributed to $P^{(d)}$.
\subsection{E1 transition: $\Psi' \rightarrow \phi_1 +\gamma,\  \phi_1
  \rightarrow J/\Psi +\gamma $}
 $\phi_1 $ is produced in the radiative decays of $\Psi'$
through the $E_1$ transition 
\begin{eqnarray}
\Psi' \rightarrow \phi_1 +\gamma,\label{radiative-decay1}
\end{eqnarray}
which is expressed by the effective interaction
\begin{eqnarray}
S_{int}=e'\int d^4x  O^{\mu\nu}  F_{\mu,\nu},  ~~O^{\mu\nu}=\epsilon^{\mu\nu\rho\sigma} \Psi'_{\mu} \phi_{1,\nu}
\label{E1-transition}
\end{eqnarray}
 in the local limit. The action Eq. \eqref{E1-transition} does  not take
 the form
 of the total derivative, but is written as
 \begin{eqnarray}
S_{int}=e'\int d^4x   \partial_{\nu}(O^{\mu\nu}A_{\mu}  )-
e'\int d^4x  (\partial_{\nu}O^{\mu\nu}) A_{\mu}. 
\label{E1-transition2}
\end{eqnarray} 
The second term shows the interaction of the local electromagnetic coupling
of the current $j^{\mu}=\partial_{\nu}O^{\mu\nu}$
and the first term shows the surface term. This leads to the constant 
 probability at a finite $T$,  $P^{(d)}$.  The induced $P^{(d)}$ for muon
 decay   was computed in 
\cite{ishikawa-tobita1}, and  large probability from $P^{(d)}$ compared
with $T\Gamma_0$ was found in the region  $cT= 1\, \text{m}$.  

The  radiative transitions of heavy
quarkonium are  deeply connected with other radiative transitions and
the detailed analysis will be presented elsewhere.

Spin $0$ and $2$ mesons , $\phi_0$ and $\phi_2$, show the same $E_1$ transitions and photons
show the same behavior from $P^{(d)}$ \cite{crystal-ball,cleo,bes,particle-data}.
A pair of  photons  of
the continuous energy spectrum  are produced in the wave zone, and are 
correlated.  On the other hand, a pair of photons of
 the discrete energy spectrum are produced in the 
particle zones and are  not correlated. 
Thus the photons  in the continuous spectrum  are different from the simple
background, and it would be possible to confirm the correlation  by 
measuring the time coincident of the two photons.

\section{Neutrino-photon interaction }
The neutrino  photon interaction of the strength   $\alpha G_F$  induced
from higher order effects vanishes due to the Landau-Yang-Gell-Mann theorem. 
But  $P^{(d)}$ is free from the theorem and gives observable effects. Moreover,
although the strength
seems  much weaker than the normal weak process, that  is enhanced
drastically if the photon's  effective mass   is extremely small. 
Because $P^{(d)}$ does not preserve Lorentz invariance, a careful treatment is
 required.
    The triangle
 diagram of Fig. 1  is expressed  in terms of the action
\begin{eqnarray}
& &S_{\nu\gamma}={1 \over 2\pi}\alpha_{em}  {G_F \over \sqrt 2}\int d^4 x f_1 {\partial \over
 \partial x_{\mu} } (J^{A}_{\mu}(x) \tilde F_{\alpha\beta} F^{\alpha\beta}),
\label{neutrino-photon}\\
& & J^{A}_{\mu}(x)=\bar \nu(x) (1-\gamma_5)\gamma_{\mu} \nu(x), \nonumber
\end{eqnarray} 
where   the axial vector meson in
 Eq. $(\ref{effective-two-photon})$ was replaced by the neutrino current. 
The mass of the neutrino is extremely small, and was  neglected. 
The  $S_{\nu\gamma}$  leads to the neutrino gamma reactions with $\Gamma_0=0, P^{(d)} \neq 0$
 on the order $\alpha G_F$.
   
\subsection{$\nu+\gamma \rightarrow \nu+\gamma$}
 The rates $\Gamma_0$  of the events 
\begin{eqnarray}
& &\nu+\bar \nu \rightarrow \gamma +\gamma \\ 
& &\nu+\gamma \rightarrow
\nu+ \gamma \nonumber
\end{eqnarray}   
 vanishes on the order $\alpha G_F$ \cite{gell-mann}  due to
Landau-Yang's theorem.   The higher order effects were 
also shown to be extremely small \cite{rosenberg} and 
these processes have been ignored. The theorem  is derived from 
the rigorous conservation law of the kinetic-energy and angular momentum.
However these do not hold in $P^{(d)}$, 
due to the interaction energy caused by the overlap of
the initial and final wave functions. Consequently, neither $P^{(d)}$
nor  the transition
probability vanish.    
 These  processes are reconsidered with  $P^{(d)}$.  

  From Eq. $(\ref{neutrino-photon})$,  we have the probability amplitude of the
  event in which  one of the photon of ${\vec
 k}_{\gamma}$   interacts with another object or  is detected 
at ${\vec X}_{\gamma}$  as   
\begin{align}
\mathcal M=-{G_F \alpha_{em} \over 2 \pi \sqrt 2}\int d^4x {\partial
 \over \partial x^{\mu}}[ \langle
 p_{\nu,2}|J_{\mu}^A(x)|p_{\nu,1} \rangle \langle ({\vec
 k}_{\gamma},{\vec X}_{\gamma},T_{\gamma})| \tilde F_{\alpha \beta}
 F^{\alpha \beta} | k_1 \rangle].
\end{align}
The amplitude is expressed in the same manner as Eq.$(\ref{amplitude-axial-vector})$,   
\begin{eqnarray}
& &\mathcal M=-i{G_F \over \sqrt 2}{2 \over \pi}\alpha N \int_{\lambda \geq 0}
 d^4x    {\partial
 \over \partial x^{\mu}}[
e^{-i(p_{\nu_1}-p_{\nu_2}+k_1)x} 
\tilde {\mathcal M}^\mu], \label{amplitude-neutrino-photon} \\
& &\tilde {\mathcal M}^\mu= f(k_1\cdot
 k_{\gamma}^2)e^{i(E(k_{\gamma}^2)(t-T_{\gamma})-{\vec k}_{\gamma}^2({\vec
 x}-{\vec X}_{\gamma}))-\chi(x) 
  }  T^{\mu},\nonumber \\
& &T^{\mu}=  \bar \nu(p_{\nu_2})
(1-\gamma_5)\gamma^{\mu}\nu(p_{\nu_1})\epsilon_{\alpha\beta\delta\zeta }
 \epsilon^{\alpha}(k^1)\epsilon^{\beta}(k_{\gamma}^2
+\delta k_{\gamma}(x) ) k_1^{\delta} (k_2+\delta k_{\gamma}(x))^{\zeta}, \nonumber\\
& &N=  \left({(2\pi)^3 \over\sigma_{\gamma} \sigma_T^2}\right)^{\frac{1}{2}} N_{\gamma}
 (2\pi)^{\frac{3}{2}}\rho_{\gamma}(k^1)\rho_{\gamma}(k_{\gamma}^2+\delta k) {1 \over
 (2\pi)^{\frac{3}{2}}} \left({m_{\nu}m_{\nu} \over E_{\nu_1}E_{\nu_2}}\right)^{\frac{1}{2}},\nonumber
\end{eqnarray}
where the corrections due to higher order in $\delta {\vec k}$ are ignored in
the following calculations as in Section (3.1). The probability averaged over the initial spin  per unit of particles of initial state,
is 
\begin{eqnarray}
& &P= {1 \over 2} \int { d\vec{{X}}_{\gamma} \over V (2\pi)^3}d{\vec k}_{\gamma}^2 
  d{\vec k}_{\nu}^2 \sum_{s_1,s_2}|\mathcal M|^2\nonumber\\
& &~~=  \int {d{\vec k}_{\gamma,2} \over
 (2\pi)^32E_{\gamma,2}}   N_0^2 (i)
(I_0 p_{\nu_1}^0(p_{\nu_1}+p_{\gamma,1}-
 p_{\gamma_2})^{0}+ I_i p_{\nu_1}^i (p_{\nu_1}+p_{\gamma_1}-
 p_{\gamma_2})^{i}) \nonumber\\
& & ~~~~~\times (4 k_1\cdot k_{\gamma} f_1)^2  
 \theta(\text{Phase-space}),\label{total-probability-gamma1} \\
& &\theta(\text{Phase-space})=\theta((p_{\nu_1}+k_{\gamma,1})^2-2(p_{\nu_1}+k_{\gamma,1})k_{\gamma,2}), \nonumber \\
& &N_0^2={1\over 2}\left({G_F \over \sqrt 2}{2 \alpha \over \pi}\right)^2 {1 \over
 4\pi}|N_{\gamma}|^2\left({1 \over \sigma_{\gamma}}\right)^3
  {1 \over 2E_{\nu}^1} {1 \over
 2E_{\gamma}^1}  \nonumber,
\end{eqnarray} 
where $V$ is the normalization volume of the initial state, and  
\begin{eqnarray}
& &f_1(k_1\cdot k_{\gamma})=
\begin{cases}
{1 \over 4k_1 \cdot k_{\gamma}};\ \text{high~energy},\\
{1 \over 2m_e^2};\ \text{low~energy},
 \end{cases}
\end{eqnarray}
where $I_0$ and $I_i$ are given in Appendix E.  
The probability $P$ is not Lorentz invariant and
the values in the CM frame of the initial neutrino and photon and those of 
 the general frame do not agree generally.

\subsubsection{Center of mass frame }

The phase space integral over the momentum  in the CM frame
\begin{eqnarray}
{\vec p}_{\nu_1}+{\vec p}_{\gamma_1}=0,\ p=|{\vec p}_{\nu_1}| 
\end{eqnarray}
is
\begin{align}
&\int {d{\vec k}_{\gamma,2} \over
 (2\pi)^32E_{\gamma,2}} (p (2 p -p_{\gamma_2}))_{0} (4 f_1 k_1\cdot k_{\gamma})^2
 \theta(\text{phase-space})\nonumber\\
&={1 \over 6  \pi^2} p^4 ;\ \text{high-energy},\\
&\int {d{\vec k}_{\gamma,2} \over
 (2\pi)^32E_{\gamma,2}} (p (2 p -p_{\gamma_2}))_{0} (4 f_1 k_1\cdot
 k_{\gamma})^2{1 \over k_{\gamma,2}^2}
 \theta(\text{phase-space})\nonumber\\
&={1 \over 12\pi^2} \left({p \over m_e}\right)^4 p^2;\text{low-energy}.
\end{align}
Thus we  have the probability 
\begin{align}
P=&
{1 \over 2}\left({G_F \over \sqrt 2}{2 \alpha \over \pi}\right)^2
\frac{1}{4\pi^{\frac{5}{2}}}
\left(\sqrt{\sigma_{\gamma}} \log(\bar \omega T)+{1 \over 2m_{\gamma}^2 \sqrt{\sigma_{\gamma}}}\right)
{p^2 \over 12};\ \text{high-energy},\\
=& {1 \over 2}\left({G_F \over \sqrt 2}{2 \alpha \over \pi}\right)^2
{1 \over 4\pi^{\frac{5}{2}}}{ 1 \over 4^3 6 \pi^{\frac{5}{2}}}
 \sigma_{\gamma}^{-\frac{1}{2}} {1 \over \epsilon}\left({p \over m_e}\right)^4
 ;\text{low-energy},  \label{neutrino-photon-smallT}
\end{align}
where  $\bar \omega$ in the log term is the average $\omega$, and
  $\epsilon$ is the deviation of the index of  refraction from unity given in
  Appendix C. The log term was ignored in the right-hand side of the low energy
  region. We found that in the majority of region, $\frac{1}{\omega E \sqrt{\sigma_\gamma}}$ becomes
  much larger than $\sqrt{\sigma_\gamma}\log(\omega T)$.  Here we assume that
  the medium is not ionized.  

\subsubsection{Moving  frame }
 At the frame ${\vec p}_{\nu_1}=(0,0, p_{\nu_1}),\ {\vec
 p}_{\gamma_1}=(0,0, -p_{\gamma_1}),\ { p}_{\nu_1} > { p}_{\gamma_1}$ 
the probability  
in  high energy region  is 
\begin{align}
P=&N_0^2  \int { d {\vec k}_{\gamma,2} \over (2\pi)^3 2E_{\gamma,2}}
 \theta((p_1+k_1)^2-2(p_1+k_1)k_{\gamma,2})\nonumber \\
& \times \biggl[(p_{\nu_1}^0(p_{\nu_1}+p_{\gamma_1}-p_{\gamma_2})^0+
 p_{\nu_1}^l(p_{\nu_1}+p_{\gamma_1}-p_{\gamma_2})^l)_{l}
(\sigma_{\gamma}^2
 \log \omega T+{\sigma_{\gamma} \over 4\omega E})\nonumber\\
&+p_{\nu_1}^i(p_{\nu_1}+p_{\gamma_1}-p_{\gamma_2})^i(\sigma_{\gamma}^2 \log
 \omega T)\biggr] \nonumber  \\
=&C_0 \left(\sqrt{\sigma_{\gamma}}\log \omega T +\frac{1}{\sqrt{\sigma_\gamma}m_{\gamma}^2}\right)
p_{\nu_1}^2; p_{\nu_1} \gg p_{\gamma_1}. \label{neutrino-photon-high}
\end{align}
In the low energy region,   the second term of $P$ is given in the form,
\begin{eqnarray}
P=C_0'   {1
 \over \sqrt{\sigma_\gamma} \epsilon }\left({ p_{\nu_1} \over
 m_e}\right)^4,
 \label{neutrino-photon-low}
 \end{eqnarray}
 and the first term  proportional to $p_{\nu_1}^6$ was ignored.
Numerical  constants $C_0$ and $C_0'$ are proportional to $({G_F \over
 \sqrt 2}{2 \alpha \over \pi})^2 $.   
 The photon effective mass at high energy region, $m_{\gamma}$,  and
the deviation of the refraction constant from unity  at low
energy region, $\epsilon$,   are
extremely small in the dilute gas, and $P^{(d)}$ becomes large in these
situations.   

\subsection{Neutrino interaction with uniform   magnetic field $\nu +B \rightarrow \nu+\gamma $}
The action Eq. $(\ref{neutrino-photon})$ leads to the coherent interactions
of neutrinos  with the macroscopic electric or magnetic fields. 
These fields   are expressed  in MKSA unit. 
Accordingly  we express the Lagrangian in MKSA unit, which is summarized inAppendix D, 
and compute the probabilities.    
The magnetic field  $B$ in the z-direction is expressed by  the field strength  
\begin{eqnarray}
F_{\mu\nu}(x)=\epsilon_{3,\mu\nu}B, 
\end{eqnarray}
and we have the action 
\begin{eqnarray}
S_{\nu\gamma}(B)= g_B  \int d^4 x  {\partial \over
 \partial x_{\mu} } (J^{A}_{\mu}(x)  F_{0,z} ),
\label{neutrino-neutrino-photon}
\end{eqnarray} 
where $g_B$ is given in Appendix D.
Because  $S_{\nu\gamma}(B)$ is reduced to the surface term, the
rate vanishes,   $\Gamma_0=0$, but $ P^{(d)} 
\neq 0$. Furthermore, $ S_{\nu\gamma}(B)$ is not  Lorentz invariant, and  
$P^{(d)}$ for $\nu_i \rightarrow \nu_j+\gamma$  becomes 
proportional to $m_{\nu}^2$ of much larger magnitude than
the naive expectation.

The amplitude is
\begin{eqnarray}
& &\mathcal M=-i N g_B  \int_{\lambda \geq 0}
 d^4x    {\partial
 \over \partial x^{\mu}}[
e^{-i(p_{\nu_1}-p_{\nu_2})x} 
\tilde {\mathcal M}^\mu], \label{amplitude-neutrino-B} \\
& &\tilde {\mathcal M}^\mu= e^{i(k_0(k_{\gamma})(x_0-X_{\gamma}^0)-{\vec k}_{\gamma}({\vec
 x}-{\vec X}_{\gamma}))-\chi(x) }  T^{\mu},\nonumber \\
& &T^{\mu}=  \bar \nu(k_{\nu_2})
(1-\gamma_5)\gamma^{\mu}\nu(k_{\nu_1})(
 \epsilon^{0}(k_{\gamma}) k_{\gamma}^z-\epsilon^{z}(k_{\gamma}) k_{\gamma}^0
 ) \nonumber,\\
& &N=  \left({(2\pi)^3 \over\sigma_{\gamma}\sigma_T^2}\right)^{\frac{1}{2}} N_{\gamma}
 \rho_{\gamma}(k_{\gamma}) {1 \over
 (2\pi)^{\frac{3}{2}}}
\left({\tilde \omega_{\nu_1}^0 \tilde \omega_{\nu_2}^0 \over k_{\nu_1}^0
 k_{\nu_2}^0}\right)^{\frac{1}{2}},
\  \rho_{\gamma}(k_{\gamma})=
\left({1 \over
 (2\pi)^32k_{\gamma}^0} {\hbar  \over \epsilon_0 }\right)^{\frac{1}{2}},\nonumber
\end{eqnarray} 
where
\begin{align}
\sum_{spin}(T^{\mu_1} T^{\mu_2*})={8 \over 2\tilde \omega_{\nu_1}^0
 2 \tilde \omega_{\nu_2}^0}(k_{\nu_1}^{\mu_1}k_{\nu_2}^{\mu_2}
 -g^{\mu_1\mu_2}k_{\nu_1}k_{\nu_2}+k_{\nu_1}^{\mu_2}k_{\nu_2}^{\mu_1})({k_{\gamma}^0}^2-{k_{\gamma}^z}^2)
. 
\end{align}
 We have the  probability from Eq. $(\ref{total-probability-gamma})$, 
\begin{eqnarray}
& &P={\tilde N_0}^2g_B^2(2\pi)^3 16 
{1 \over 2E_{\nu_1}} {1 \over \epsilon_0} 
\int {d{\vec k}_{\gamma} \over (2\pi)^3 2E_{\gamma}} ({k_{\gamma}^0}^2-{k_{\gamma}^3}^2) \nonumber\\ 
& &\times  (i)(I_0  k_{\nu_1}^0 (k_{\nu_1}-k_{\gamma})^0 +I_l k_{\nu_1}^l (k_{\nu_1}-k_{\gamma})^l) 
\theta(\text{phase-space}), \label{probabiiity-B}\\
& & \theta(\text{phase-space})=\theta(\delta
 \omega_{\nu_{12}}^2-2k_{\nu_1}k_{\gamma}) 
 \nonumber,  
\end{eqnarray}
where ${\tilde N_0}^2=(\pi \sigma_{\gamma})^{-\frac{3}{2}}$, $\delta
\omega_{12}^2=(\tilde \omega_{\nu_1}^0)^2-(\tilde \omega_{\nu_2}^0)^2 $, $I_{0,l} $ and $I_{T,i}$
are given in Appendix E.

The convergence condition on the light-cone singularity is satisfied 
 in the kinematical region, $\theta(\text{phase-space})$. Thus
the momentum satisfies 
\begin{eqnarray}
2(k_{\nu_1}^0k_{\gamma}^0-k_{\nu_1} k_{\gamma} \cos \theta) \leq {\delta 
\omega_{12}}^2.
\end{eqnarray}
Solving $k_{\gamma}$, we have the condition for  fraction 
$x={k_{\gamma} \over k_{\nu_1}}$,
\begin{eqnarray}
& &\alpha_{-}  \leq x \leq \alpha_{+},\\
& &\alpha_{\pm}={\delta \omega_{12}^2 \pm \sqrt{\delta
 \omega_{12}^4-4(\tilde \omega_{\nu_1}^0)^2 m_{\gamma}^2 } \over
 2(\tilde \omega_{\nu_1}^0)^2}, \label{fractios}
\end{eqnarray}
where $\alpha_{\pm}=O(1)$. 
The process $\nu_i \rightarrow \nu_j+\gamma$ occurs with  the  probability Eq. $(
\ref{probabiiity-B} )$ if
\begin{eqnarray}
 m_{\gamma}^2 \leq  {({\delta \omega_{12}^2})^2 \over 4({\tilde \omega_{\nu_1}^0})^2}, \label{mass-inequalitye}
\end{eqnarray}
which is satisfied   in dilute gas. If the inequality
Eq. $(\ref{mass-inequalitye} )$ is not satisfied, this  probability vanishes.

The probablity $P$ reflects the large overlap of initial and final states and is not 
Lorentz invariant. 
Consequently,  although the integration region in Eq. $(\ref{probabiiity-B})$ is  narrow in phase space determined 
by  $\theta(\text{phase-space})$, which is proportional to the
mass-squared difference, $\delta
\omega_{12}^2$, the integrand is as large as
$p_{\nu}^3$.  $P$ becomes much larger than the value
obtained from Fermi's golden rule. 

\subsubsection{High energy  neutrino}
At high energy, Eq. $( \ref{high-energy-I})$ are substituted.  For   the case that ${\vec p}_{\nu_1}
$ is  non-parallel 
to ${\vec B}$, we have
\begin{eqnarray}
& &{d P \over d x}={\tilde N_0}^2 g_B^2 (2\pi)^3 16 {1 \over
 \epsilon_0} (1-\cos^2 \zeta)  (i)I_0 k_{\nu_1}^3 C(m,x) \label{non-parallel-B}\\
& &C(m,x)= m_{\gamma}^2x^4 -(\delta \omega_{12}^2+m_{\gamma}^2) x^3 +
(\tilde \omega_{\nu_1})^2- \omega_{12}^2)x^2 - ({\tilde \omega_{\nu_1}^0})^2 x , \nonumber 
\end{eqnarray}
where $\zeta$ is the angle between  ${\vec p}_{\nu_1}$ and ${\vec B}$,
\begin{eqnarray}
{\vec B}\cdot{\vec p}_{\nu_1}=Bp_{\nu_1} \cos \zeta.
\end{eqnarray}

The integral $I_0$ is almost independent of $k_{\gamma}$. Ignoring the dependence,  
we integrate the photon's momentum, and  have
\begin{eqnarray}
& &P={\tilde N_0}^2 g_B^2 (2\pi)^3 16 {1 \over
 \epsilon_0} (1-\cos^2 \zeta)  (i)I_0 k_{\nu_1}^3 C(m) \label{non-parallel-Btotal}\\
& &C(m)= {1 \over 5}m_{\gamma}^2(\alpha_{+}^5-\alpha_{-}^5)-{1 \over
 4}(\delta \omega_{12}^2+m_{\gamma}^2) (\alpha_{+}^4-\alpha_{-}^4)
\nonumber \\
& &+{1 \over 3}(({\tilde \omega_{\nu_1}^0})^2- \omega_{12}^2)(\alpha_{+}^3-\alpha_{-}^3)-{1 \over
 2}({\tilde \omega_{\nu_1}^0})^2(\alpha_{+}^2-\alpha_{-}^2). \nonumber
\end{eqnarray}
The probability $P$ of Eq. $(\ref{non-parallel-Btotal})$ is proportional to   $({\tilde
\omega_{\nu}^0})^2k_{\nu_1}^3$, which is very
different from the rate  of  the normal neutrino radiative decay, 
$ \Gamma = G_F^2 m_{\nu}^5 ({m_{\nu_1}/E_{\nu_1}})\times (\text {numerial
 ~factor})$, especially for high energy neutrino. Moreover, $I_0 ={1 \over m_{\gamma}^2}$, 
 can be  extremely large in dilute gas, thus $P$ is  enormously
enhanced.  

If  the momentum of initial neutrino is parallel to the magnetic field,  
$\zeta =0$, we have 
\begin{eqnarray}
& &P={\tilde N_0}^2 g_B^2 (2\pi)^3 16 {1 \over
 \epsilon_0}   (i)I_0  k_{\nu_1}D(m), \label{parallel}\\
& &D(m)= \delta
 \omega_{12}^4 {\alpha_{+}^2-\alpha_{-}^2 \over 2}-\delta \omega_{12}^2
({\tilde \omega_{\nu_1}^0})^2 (\alpha_{+}-\alpha_{-}-2 {\alpha_{+}^2 -\alpha_{-}^2 \over
 2} ) \nonumber \\
& &~~~~~~~~+({\tilde \omega_{\nu_1}^0})^2(\log{\alpha_{+} \over
 \alpha_{-}}-\alpha_{+}+\alpha_{-}). \nonumber
\end{eqnarray} 
$P$  in Eq. $( \ref{parallel} )$ is proportional to  
$({\tilde \omega_{\nu}^0})^4k_{\nu_1}$, and  is negligibly 
small compared to that of  Eq. $(\ref{non-parallel-Btotal})$.    
\subsubsection{ Low energy neutrino }

At  low energy, Eq. $ (\ref{low-energy-I})$ are substituted. $I_0$
is inversely proportional to $k_{\gamma}^2$ and we have
\begin{eqnarray}
& &{d P \over dx}={\tilde N_0}^2 g_B^2 4\pi   E_{\nu_1} {1
 \over \epsilon_0} (1 -\cos^2 \zeta) {1 \over \epsilon}  (1-x)[\delta \omega_{12}^2-xm_{\gamma}^2-{1
 \over x}m_{\gamma}^2], \nonumber \label{low energy-magnetic-field1} 
\end{eqnarray}
and 
\begin{eqnarray}
& &P={\tilde N_0}^2 g_B^2 4\pi   
{ 1 \over \epsilon_0} (1-\cos^2 \zeta) k_{\nu_1} {1 \over \epsilon}C_{low}(m),\\ 
& & C_{low}(m)=
\delta \omega_{12}^2 (\alpha_{+}-\alpha_{-}-{\alpha_{+}^2 -\alpha_{-}^2
\over 2})-m_{\gamma}^2
({\alpha_{+}^2-\alpha_{-}^2 \over 2}
-{\alpha_{+}^3-\alpha_{-}^3 \over 3} )\nonumber \\
& &~~~~~+
(\tilde{\omega}_{\nu_1}^0)^2
(\alpha_{+}-\alpha_{-})
-({\tilde \omega_{\nu_1}^0})^2 log{\alpha_{+} \over \alpha_{-}}. 
\nonumber
\end{eqnarray}
Using $\alpha_{\pm}$,  $C(m)$ and $D(m)$ are computed easily.

\subsection{Neutrino interaction with uniform  electric  field $\nu +E \rightarrow \nu+\gamma $}
For a uniform electric field in the z-direction,
\begin{eqnarray}
F_{\mu\nu}(x)={E \over c}  \epsilon^{03\mu\nu},
\end{eqnarray}
we have
\begin{eqnarray}
& &T^{\mu}=  \bar \nu(k_{\nu_2})
(1-\gamma_5)\gamma^{\mu}\nu(k_{\nu_1})(
 \epsilon^{x}(k_{\gamma}) k_{\gamma}^y-\epsilon^{y}(k_{\gamma}) k_{\gamma}^x
 ) ,\label{neutrino-electric-field}\\
& &\sum_{spin}(T^{\mu_1} T^{\mu_2*})=8{1 \over 2\omega_{\nu_1}^0
 2\omega_{\nu_2}^0}(k_{\nu_1}^{\mu_1}k_{\nu_2}^{\mu_2}
 -g^{\mu_1\mu_2}k_{\nu_1}k_{\nu_2}+k_{\nu_1}^{\mu_2}k_{\nu_2}^{\mu_1})\nonumber \\
& &\times ({k_{\gamma}^x}^2+{k_{\gamma}^y}^2).
\end{eqnarray}
 $\nu_i \rightarrow \nu_j+\gamma$ in the electric field is almost the
 same as that in the magnetic field.
\subsubsection{High energy  neutrino }
We have the probability in the high energy region,  
\begin{eqnarray}
& &P={\tilde N_0}^2 g_E^2 (2\pi)^3 16 {1 \over
c^2 \epsilon_0} (1-\cos^2 \zeta) (i) I_0 k_{\nu_1}^3 C(m) \label{non-parallel-total},\\
& &C(m)= {1 \over 5}m_{\gamma}^2(\alpha_{+}^5-\alpha_{-}^5)-{1 \over
 4}(\delta \omega_{12}^2+m_{\gamma}^2) (\alpha_{+}^4-\alpha_{-}^4)
\nonumber \\
& &+{1 \over 3}(({\tilde \omega_{\nu_1}^0})^2- \omega_{12}^2)(\alpha_{+}^3-\alpha_{-}^3)-{1 \over
 2}({\tilde \omega _{\nu_1}^0})^2(\alpha_{+}^2-\alpha_{-}^2), \nonumber
\end{eqnarray}
where
\begin{eqnarray}
{\vec E}{\vec k}_{\nu_1}=Ek_{\nu_1} \cos \zeta.
\end{eqnarray}
\subsubsection{Low energy neutrino }
The probability    in the low energy region is
\begin{eqnarray}
& &P={\tilde N_0}^2 g_E^2 4\pi   
{ 1 \over c^2 \epsilon_0} (1-\cos^2 \zeta) k_{\nu_1} {1 \over \epsilon}C_{low}(m),\label{non-parallel-low-E}\\ \nonumber
& & C_{low}(m)=
\delta \omega_{12}^2 
\left(\alpha_{+}-\alpha_{-}-{\alpha_{+}^2 -\alpha_{-}^2
\over 2}\right)
-m_{\gamma}^2
\left({\alpha_{+}^2-\alpha_{-}^2 \over 2}
-{\alpha_{+}^3-\alpha_{-}^3 \over 3} \right) \\
& &~~~~~+({\tilde \omega_{\nu_1}^0})^2(\alpha_{+}-\alpha_{-})
-({\tilde \omega_{\nu_1}^0})^2 log{\alpha_{+} \over \alpha_{-}}. 
\nonumber
\end{eqnarray}

 \subsection{Neutrino interaction with nucleus electric field 
$\nu +E_{nuc} \rightarrow \nu+\gamma $}
In space-time near a nucleus,  there is  the Coulombic   electric  
field $E_{nucl}$ due to the nucleus, and one 
of $F_{\mu \nu}$ in the action is replaced with  $E_{nucl}$. The rate 
estimated in Ref. \cite{rosenberg} was much smaller by a factor $10^{-4}$ or
more than the value of the
normal process due to charged current interaction. Here we estimate
$P^{(d)}$ for the same process. The
action becomes  
\begin{eqnarray}
S_{\nu\gamma}(E_{nucl})={\frac{1}{2\pi}}\alpha_{em}  {G_F \over \sqrt 2}  
\int d^4 x f_1 {\partial \over
 \partial x_{\mu} } (J^{A}_{\mu}(x)  F_{\mu \nu} F_{nucl}^{\mu \nu}),
\label{neutrino-neutrino-photon1}
\end{eqnarray} 
which  causes the unusual radiative interaction  of neutrino in matter. 
The probability $P^{(d)}$ of  the high energy neutrino, where  $2 k_1\cdot k_2 \gg m_e^2$,
and  we substitute the value $f_1={ 1 \over 8m_l^2}$. Since $F_{nucl}^{\mu\nu}$ 
due to a bound nucleus is short-range, the probability is not enhanced. 
\subsection{Neutrino interaction with laser  wave  
$\nu +E_{laser} \rightarrow \nu+\gamma $}
In the scattering of neutrino with a classical electromagnetic 
wave due to laser,   one 
of $F_{\mu \nu}$ in the action is replaced with the electromagnetic
field  $E_{laser}$ of laser of the form
\begin{eqnarray}
F_{laser}^{\mu\nu} =E^i \epsilon^{0i\mu\nu}e^{ik_1 x}.
\end{eqnarray}
The probability $P^{(d)}$ of this  process is computed with the
action
\begin{eqnarray}
S_{\nu\gamma}(E_{nucl})={1 \over 2\pi}\alpha_{em}  {G_F \over \sqrt 2}  \int d^4 x f_1 {\partial \over
 \partial x_{\mu} } (J^{A}_{\mu}(x)  F_{\mu \nu} F_{laser}^{\mu \nu}),
\label{neutrino-neutrino-photon2}
\end{eqnarray} 
where  $2 k_1\cdot k_2 \approx 0 $,
and  we substitute the value $f_1={ 1 \over 8m_l^2}$. 
The amplitude and probability are almost equivalent to those of
the uniform electric field.

\section{Implications  to neutrino reactions in matter and  fields  } 
  An initial neutrino  is
  transformed to another  neutrino and a photon following the  probability
  $P^{(d)}$. The photon 
in the final state interacts with a microscopic object  in matter with 
the electromagnetic  interaction, and loses the energy.
 Thus the size  $\sigma_{\gamma}$  in $P^{(d)}$
is  determined by its wave function, and the 
  probability   $P^{(d)}\times \sigma_{\gamma A}$,
  where $\sigma_{\gamma A}$   is the cross section of the  photon and   
  is much larger than that of  weak reactions, determines the effective
  cross section of the whole process.  Hence the effective
  cross section  can be as large
  as that of the normal weak process caused by the charged current
  interaction.  

\subsection{Effective cross section}
 
The probability of the
 event that the photon reacts on another object is expressed by  
$P^{(d)}$ in  Eqs. $(\ref{neutrino-photon-smallT})$, $(\ref{neutrino-photon-high})$,
 and $(\ref{neutrino-photon-low})$, and that of  the final photon. 
  If  a system initially has  photons of  density
 $n_{\gamma}(E_{\gamma})$, the number of 
photons are multiplied, and the  probability of the event that 
 the initial neutrino is transformed 
  is given by  $P^{(d)}\times n_{\gamma} $. In the system of electric or
  magnetic fields,  
  the initial neutrino is transformed to the final  neutrino and
  photon. 
Hence  $P^{(d)}\times n_{\gamma} $ for the former case, and $P^{(d)}$
for the latter case are important parameters to be  compared with the experiments.

The effective cross section, for the process where  the photon in the 
final state  interacts   with atoms $A$  of the cross section   
$\sigma_{\gamma  A}$,   is 
\begin{eqnarray}
\sigma^{(d)}_{\gamma A}=P^{(d)}(\gamma) n_{\gamma} \times \sigma_{\gamma A},
\label{effective-cross1}
\end{eqnarray}
for the former case, and 
\begin{eqnarray}
\sigma^{(d)}_{\gamma A}=P^{(d)}(\gamma)  \times \sigma_{\gamma A}.
\label{effective-cross2}
\end{eqnarray}
for the latter cases.

The cross sections  Eqs. $(\ref{effective-cross1})$ and 
$(\ref{effective-cross2}) $ 
 are  compared with that of the charged current weak process  
\begin{eqnarray}
\sigma_{\nu A}^{weak}={{G_F^2} \over 2}E_{\nu}M_A.
\label{weak-cross} 
\end{eqnarray}
Since  $ \sigma_{\gamma A}$ is much 
larger than 
$\sigma_{\nu A}^{weak}$,  by a factor $10^{14}$ or more, $\sigma^{(d)}_{\gamma A} $ in
 Eqs. $(\ref{effective-cross1})$ and $(\ref{effective-cross2})$  can be as
large  Eq. $(\ref{weak-cross} )$, if $P^{(d)}$ is around
$10^{-14}$. Accordingly,  $10^{-14}$ or  $10^{-15}$ is the critical value
 for the photon neutrino process   to be relevant and
important.  If the value is larger, then the   reaction that  is dectated to vanish 
due to Landau-Yang's theorem  manifests with a sizable
probability.

\subsection{ $\nu+\gamma
  \rightarrow  \nu  + \gamma $}
  The  probability $P^{(d)}$ is of the order of $\alpha G_F$  
and almost independent of time. The probability in this order 
has been  considered  vanishing, and this process has not been studied.
If the magnitude  is sizable,   these neutrino processes  should be included   in
  astronomy and others.     The process
$\nu+\bar \nu \rightarrow \gamma +\gamma$ is almost equivalent to   
$\nu+\gamma  \rightarrow  \nu  + \gamma $, and  we do not 
study in this paper.

A system of high temperature has  many photons, and 
a neutrino makes a transition through its collision with the photons.  
 The probability is determined by the   product  between the number of
 photons $n_{\gamma}$ and each probability
\begin{eqnarray} 
P^{(d)}(\gamma) n_{\gamma}. 
 \end{eqnarray}
 In a thermal equilibrium of higher temperature, the
density is about
\begin{eqnarray}
n_{\gamma} =(kT)^3, 
\end{eqnarray}
 and we have the product for a head-on collision 
\begin{eqnarray*}
& & P n_{\gamma}=(G_F\alpha)^2{2 \over 8 \pi^{9/2} m_{\gamma}^2
  \sigma^{1/2}}{p_{\nu}^2 \over 12} (kT)^3
\end{eqnarray*}

\subsubsection{The sun } 

In the core of the sun,
\begin{eqnarray}
& & R=10^9 \text{ meters}, \\
& & kT \approx 2 \text{ keV},\nonumber
 \end{eqnarray}
 and the
solar neutrino has the  energy around  $1-10$ MeV. The
 photon's energy distribution is given by the Planck distribution, and 
   the mean free path  for the head-on collision is  
\begin{eqnarray}
l={ 1 \over P^{(d)}(\gamma) n_{\gamma} \sigma_{\gamma A}}= 5\times 10^{15}\text{meters},
\end{eqnarray}
where $\sigma_{\gamma A}=10^{-24}{\text{cm}}^2$, $m_\gamma = 1$ eV and $n_A=10^{29}/{\text{cm}}^3$
are used. The value is much longer than the sun's radius.

For the neutrino of   higher energy, we use Eq. ($\ref{neutrino-photon-high}$), and have
\begin{eqnarray}
& &l={1 \over P^{(d)}(\gamma) n_{\gamma}n_A}=6.2 \times 10^{9} \left({p_{\nu}^0
 \over p_{\nu}}\right)^2 \text {meters}, \\
& & p_{\nu}^0=10 \text{ GeV}, \nonumber
\end{eqnarray}
thus the length exceeds the sun's radius for $p_{\nu} >25{\text{ GeV}}$. 
The high energy neutrino does not escape from the core  if the energy is
higher than around $25$ GeV. 

\subsubsection{  Supernova}

In supernova, the temperature is as high as  $10$ MeV and the probability
becomes much higher. We have
\begin{eqnarray}
& &l={ 1 \over P^{(d)}(\gamma) n_{\gamma} n_A}=1 {\text m}   , \\
& &~~{p_{\gamma} \over m_{\gamma}}=10^7,\ n_A=10^{25}/{\text cm}^3 \nonumber
\end{eqnarray}
or 
\begin{eqnarray}
& &l=10^4 {\text m}   , \\
& &~~{p_{\gamma} \over m_{\gamma}}=10^5,\ n_A=10^{24}/{\text cm}^3. \nonumber
\end{eqnarray}
 The mean free path 
 becomes smaller in the lower matter density region.
Thus in the region of small photon's effective mass,  the neutrino does
 not escape but loses 
its majority of  energy. This is totally different
from the standard behavior of the supernova neutrino.   
  
\subsubsection{ Neutron star}

If magnetic field is as high as  $10^9$ [Tesla],
then the probability becomes large. The energy of the neutrino is
 transfered to the photon's energy.

\subsubsection{Low energy reaction}
In low energy region,
$P^{(d)}$ behaves as Eq.$(\ref{neutrino-photon-smallT})$ and 
\begin{eqnarray}
\sigma_{\gamma A} \rightarrow C E_{photon}^{-3.5} , E_{photon}
 \rightarrow 0.
\end{eqnarray}
The effective transition probability $P^{(d)} n_{\gamma}$ and the cross
section depend on the
photon density.

\subsection{$\nu +(E,B) ~~\rightarrow \nu'+\gamma$}
The radiative transition of one neutrino  to another lighter neutrino  
and photon in the electromagnetic field  occurs with the probability  $P^{(d)}$. 
Because $P^{(d)}$ is not proportional to $T$ but almost constant, 
the number of parents decreases fast at small $T$, and remains the same afterward without decreasing. 
 Now the photon in the final state reacts with matter with sizable
 magnitude and the  probability of whole process is
expressed with the effective cross section. 

\subsubsection{High energy neutrinos }
The transition probability of the high energy neutrino in the magnetic 
field $B$ [Tesla] and electric field $E$ [V/meter]  are
\begin{eqnarray}
& &P_B=4\alpha\left({ecB \over m_e c^2}\right)^2{G_F^2 \over 2}\pi^{-3/2}{1 \over
 4m_{\gamma}^2\sqrt{\sigma_{\gamma}}}p_{\nu}^3C(m_{\nu}){1 \over
 m_e^2}, \\
& &P_E=4\alpha\left({eE \over m_e c^2}\right)^2{G_F^2 \over 2}\pi^{-3/2}{1 \over
 4m_{\gamma}^2\sqrt{\sigma_{\gamma}}}p_{\nu}^3C(m_{\nu}){1 \over
 m_e^2}. 
\end{eqnarray}
For the parameters,
\begin{eqnarray}
m_{\gamma}c^2=10^{-9}\text{ eV},\ \sqrt{\sigma_{\gamma}}=10^{-13} \text{ meters},
\end{eqnarray}
\begin{figure}
\centering{\includegraphics[angle=-90,scale=0.45]{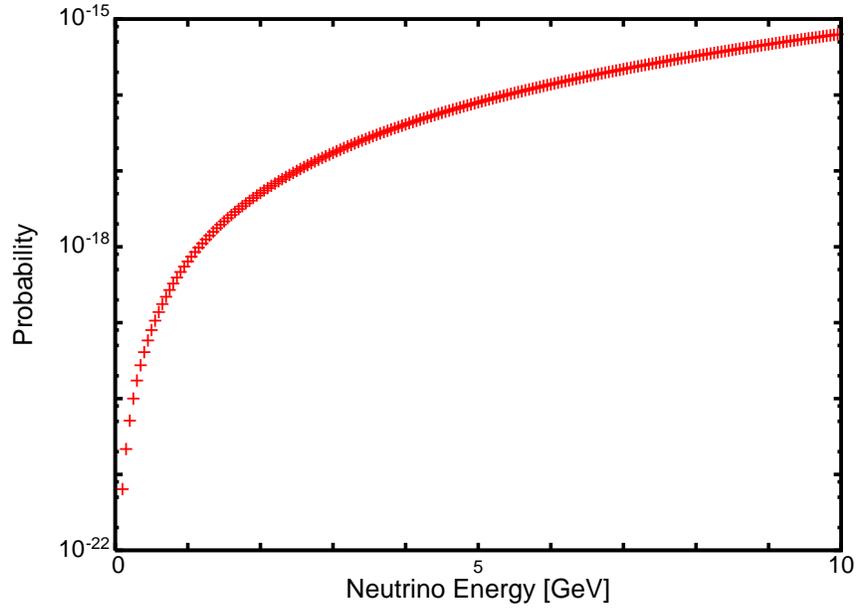}
\caption{$E_\nu$ dependence of the probability Eq. \eqref{eq-neu-B-prob} is shown. $B=10$ Tesla is used for the calculation.}
}
\end{figure}
they become  in the magnetic field  
\begin{eqnarray}
& &P_B=6.4\times 10^{-27} \left({B \over B_0}\right)^2\left({p_{\nu} \over
 p_{\nu}^{(0)}}\right)^3 \label{eq-neu-B-prob},\\
& & p_{\nu}^{(0)}=10\text{ MeV},\ B_0=1 \text{ Tesla}. \nonumber
\end{eqnarray}
and in the electric field
\begin{figure}
\centering{\includegraphics[angle=-90,scale=0.45]{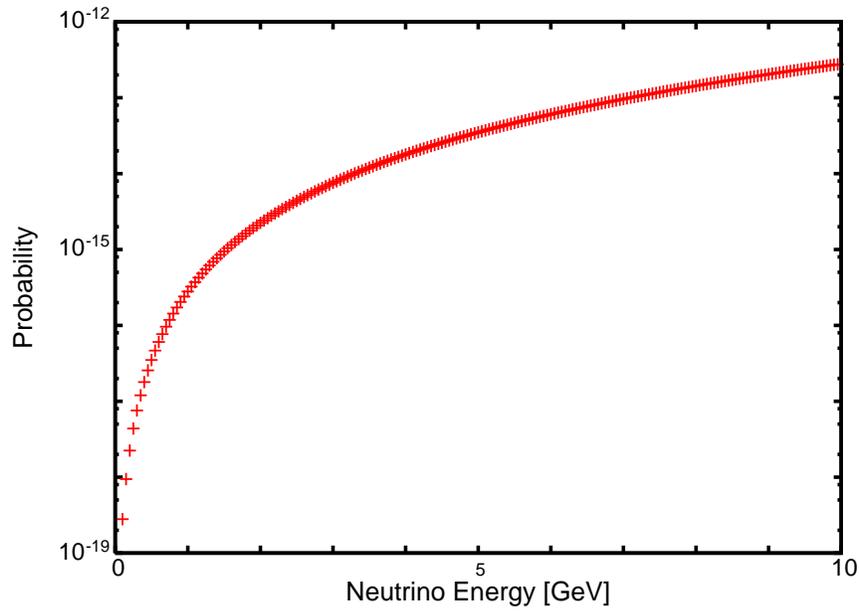}
\caption{$E_\nu$ dependence of the probability Eq. \eqref{high2-E-P} is shown. $E=100$ GV/m is used for the calculation.}
}
\end{figure}
\begin{eqnarray}
& &~~P_E=2.8\times 10^{-20} \left({E \over E_0}\right)^2\left({p_{\nu} \over
 p_{\nu}^{(0)}}\right)^3, \label{high2-E-P} \\
& &
 p_{\nu}^{(0)}=10\text{ MeV},\  E_0=10^3 \text{ GV/m} ,\nonumber
\end{eqnarray}
Neutrino energy dependences are shown in Figs. 3 ($\nu + B$) and 4 ($\nu + E$).

 \subsubsection{Low energy neutrinos }
The transition probability of the low  energy neutrino are
\begin{eqnarray}
& &P_B=4\alpha\left({ecB \over m_e c^2}\right)^2{G_F^2 \over 2}\pi^{-3/2}{1 \over
 \epsilon \sqrt{\sigma_{\gamma}}}p_{\nu} C(m_{\nu}){1 \over
 m_e^2} \\
& &P_E=4\alpha\left({eE \over m_e c^2}\right)^2{G_F^2 \over 2}\pi^{-3/2}{1 \over
 \epsilon \sqrt{\sigma_{\gamma}}}p_{\nu} C(m_{\nu}){1 \over
 m_e^2}. 
\end{eqnarray}
They become  in the typical situations 
\begin{eqnarray}
& &\epsilon =10^{-20},\ \sqrt{\sigma_{\gamma}}=10^{-13} \text{ m},
 p_{\nu}=1 \text{ eV},\ E_0=10^3 \text{ GV/m},  \label{low1-E-P} \\
& &~~P_E=6.4\times 10^{-38} \left({E \over E_0}\right)^2.\nonumber
\end{eqnarray}

The probabilities Eqs. $(\ref{eq-neu-B-prob}),\ (\ref{high2-E-P}), \text{ and }(\ref{low1-E-P})$ are caused by the overlap of waves on the parent and
 daughters. The phases of waves become cancelled at small   $m_{\gamma}$
 or $\epsilon$, and  more waves are added constructively then. Thus 
the effect become larger  as they become  smaller. The probability 
shows this  and  is  inversely proportional to $m_{\gamma}^2$ or
 $\epsilon$. They  become  extremely large as $m_{\gamma} \rightarrow 0$ or
 $\epsilon \rightarrow 0$. 
\subsection{Table of  processes}
The  $P^{(d)}$ could be  tested in various neutrino processes of wide
energy regions. 
\begin{table}[h]
 \begin{tabular}{|c|c|c|c|c|}\hline
  Neutrino source& Energy& Flux& B& E\\\hline
Solar neutrino &1--10 MeV& $10^{15}/ { {\text m}^2 \text s}$& 1--5 Tesla & --\\\hline
Reactor neutrino & 3--8 MeV & $10^{20}/{\text s}$ & 1--5 Tesla & -- \\\hline
Cumulonimbus cloud & 1--10 MeV & $10^{15}/{ {\text m}^2 \text s}$(solar $\nu$) & $10^{-5}$ Tesla & $10^3$ GV/m\\\hline
Accelerator neutrino & 0.5--50 GeV & $10^{20}$& 1--5 Tesla & --\\\hline
Cosmic neutrino & $\sim 10^6$ GeV & $10^{-9}$ GeV/cm${}^{2}$s sr& $10^{-5}$ Tesla & --\\\hline
Relic neutrino &$1$ meV & $10^{20}/\text{year}$ & -- &$10^5$ GV/m\\\hline
 \end{tabular}
\end{table}
 
The diffractive term $P^{(d)}$  becomes substantial in magnitude   at  
high energy or fields. So this process may be relevant to the neutrino
of high energy or at high fields, and   may give new insights into or
measurability of  the following processes.

1. The accelerator neutrino has high energy and total intensity of the
 order $10^{20}$. At $B=10$ Tesla and $P_{\nu}=1(10)$ GeV, we have $P \approx
 10^{-18}(10^{-15})$, and using the laser of $E_0=10^3\text{ GV/m}$ 
at $P_{\nu}=1(10)\text{ GeV}$, we
 have $P \approx 10^{-16}(10^{-13})$. The neutrino from the accelerator may be
 probed by detecting the final gamma rays. For example, at the beam dump
 of LHC we may set up laser to detect the neutrino interaction.

 2. For the reactor neutrino, of the flux  $\approx 10^{20}/$s per reactor,
  a detector  of high magnetic field of the order $10$ Tesla, or of
 high intensity laser, may be able to  detect the neutrino.

3. Direct observations of solar  neutrinos, which  have  the energy 
in $0.5$ MeV to $10$ MeV   and flux around $10^{15}$/{m${}^2$ sec}, using
the detector with strong magnetic field similar to that for axion search
\cite{axion} would be possible.

 4. In Supernova or neutron star, the neutrino photon reaction  would give
a new important process, because the final photon interacts with matter
 strongly. As to the detectability of neutrinos from the sun, $10$ GeV is
 the threshold from the sun, while they from SN interact with $10$ MeV photons.
   
  5.  We find
   that  $P^{(d)}$  becomes maximum at $\zeta={\pi \over 2}$  from   
Eq.$(\ref{non-parallel-low-E})$, and may apply this effect to enhancing the
   neutrino flux.   The term also has the momentum dependence, which may
   be exploites for ``optical'' effect of neutrino through the photon interaction.

   6. The photon neutrino reaction may be useful for the relic neutrino
   detection. The reaction rate may be enhanced with such method as the
   neutrinos mirrors that collect them.\cite{relic-neutrino} We may take
   advantage of the above effect.

 7. The probability becomes huge at extremely high energy. So, this
 process may be  relevant to ultra-high energy neutrino process. 
\cite{ice-cube,auger,telescope-array}),  

 8.  Neutrino may interact  with electromagnetic fields in Cumulonimbus cloud. 

8-1. Lightening  has  total energy of the order
$900 \text{ MJ} \approx 10^{9} $ CV and current $10^{6}$ A in a short period. 
$cB$  at a radius $r=1\text{ cm}$ is $1.5\times 10^{9} N/C$. Assuming
$E=cB=1.5\times 10^{9}$ N/C, and $m_{\gamma}c^2=10^{-11}$ eV, $P_{\nu}=10$ MeV
 we have  $P_{E+B} \approx 10^{-15} $. The neutrino  inevitably  
loses its energy, and the photons of the continuous spectrum are
emitted. This may be related with the
upper-atmospheric  lightening \cite{upper-lightening} .

 8-2. The gamma rays observed  in Cumulonimbus cloud, 
 \cite{gamma-cumulonimbus },  may be connected with 
the diffractive component.

9. Primordial magnetic fluctuations with zero frequency
\cite{promordial-magnetic-field} may have interacted with neutrinos
before neutrino detachment from the hot neutrino plasma($>$GeV
temperature) during the Big Bang. Such signatures may be carried by
neutrinos (which is now relic neutrinos).
 
\section{Summary and the future prospect}

We  have found  that the photon interaction expressed by the total derivative 
 Eqs. $(\ref{effective-interaction})$ or $(\ref{neutrino-photon})$, which are derived from the triangle diagram in the
 standard model, causes the unusual  transitions characterized  by  the
 time-independent  probability.  The interaction Lagrangian of this form
does not give rise to any physical effect  in classical physics,
because the equation of motion is not affected. In quantum mechanics,
 this assertion is correct for the transition rate $\Gamma_0$. However,
 this does not apply to 
 the diffractive term $P^{(d)}$, which manifests the wave characteristics of
 the initial and final waves.  Our results show  that $P^{(d)}$  is relevant
 to experiments and important in understanding many phenomena in nature. 
 
The neutral particles do not interact with the photon in classical
 mechanics. In quantum field theory,
 the vacuum fluctuation expressed by triangle diagram gives
the effective interaction to the  neutral
particles such as  $1^{+}  \text{ meson} \rightarrow \gamma \gamma$ and
 $\nu+\gamma (B,E)\rightarrow \nu+\gamma$. However, they  have vanishing
 rates due to Landau-Yang-Gell-Mann's theorem. $P^{(d)}$ 
 does not vanish, nevertheless,   and holds unusual properties  such as the
 violation of the kinetic-energy conservation and that of Lorentz
 invariance. Furthermore the magnitudes  become comparable to or even
 larger than the
 normal weak processes.  Accordingly, the two photon or two gluon decays of the
 neutral axial vector mesons composed of a pair of electrons or
 quarks come to have finite decay probabilities. They will be tested
 in experiments. The neutrino photon processes, which  have been
 ignored, also have finite probabilities from  $P^{(d)}$. It will be
 interesting to observe the neutrino
 photon processes directly using electric or   magnetic fields, or laser
 and neutrino beams  in various energy regions.
  The diffractive probability $P^{(d)}$ would be also important for  understanding the
 wide neutrino processes in earth, star, astronomy, and cosmology.      

 The diffractive probability $P^{(d)}$ is caused by the overlap of wave
 functions of the parent and decay products, which makes the interaction
 energy finite and   the  kinetic energy  vary. Consequently the final
 state has continuous spectrum  of the kinetic energy and  possesses 
a wave nature unique to the waves.
The unique feature of  $P^{(d)}$, i.e., independence of
on the time-interval $T$, shows that the number of parents,
which decrease like $e^{-\Gamma t}$ in the normal decay, is constant now.
The state of parent and daughter  is 
expressed by the quasi-stationary
states, which is expressed by  the superposition of different energies
 and different from the normal stationary state  of the form
$e^{-i Et \over \hbar} \psi(x_i)$.

The  
probability of the events that  the neutrinos or photons are detected 
  is   computed  with $S[T]$ that satisfies the boundary condition
  of the physical processes.
  Applying 
$S[T]$  we
obtained the results that can be compared with experiments.   
The 
 pattern of the probability  is determined by the   
difference of angular velocities,
$\omega =\omega^E-\omega^{dB}$, 
where $\omega^E={E_\nu/\hbar}$ and
$\omega^{dB}={c|\vec{p}_\nu|/\hbar}$. 
The quantity $\omega $ takes 
the  extremely small  value $  {m^2c^4}/{(2E \hbar)}$ for light
 particles such as  neutrinos or photons \cite{particle-data} in matter.
Consequently, the  diffraction term  becomes finite in the  macroscopic 
spatial region of $r \leq \frac{2\pi E \hbar c}{
m^2c^4}$. 

This allows us to introduce a new class of experimental
 measurement possibilities of  the deployment of photons to detect
 weakly interacting  particles such as neutrinos. Because the modern
 technology on the electric and magnetic fields  and  laser, a large
 number of coherent photons  are possible and the effects we have
 derived may have important implications in detecting and enhancing the
 measurement of neutrinos with photons. We see a variety of detectability
 opportunities that have eluded   attention till now. These happen
 either with high energy neutrinos such as from cosmic rays and
 accelerators; or in high fields (such as intense laser and strong
 magnetic fields ). In the latter examples, neutrinos from reactors,
 accelerators, the sun, supernovae, thunder clouds and even polar ice
 may be  detected with enhanced probabilities using intense lasers. We
 also mention the probability estimate for the primordial relic
 neutrinos and embedded information in them.

\section*{Acknowledgments}
 This work was partially supported by a 
Grant-in-Aid for Scientific Research (Grant No. 24340043). The authors  
thanks Dr. Atsuto Suzuki, Dr. Koichiro Nishikawa, Dr. Takashi Kobayashi,
Dr. Takasumi Maruyama, 
Dr. Tsuyoshi Nakaya, and Dr. Fumihiko Suekane,
 for
useful discussions on 
the neutrino  experiments and Dr. Shoji Asai, Dr. Tomio Kobayashi,
Dr. Toshinori Mori, Dr. Sakue Yamada, Dr. Kensuke Homma, Dr. Masashi
Hazumi, Dr. Kev Abazajian, Dr. S. Barwick,
Dr. H. Sobel, Dr. M. C. Chen, Dr. P. S. Chen, Dr. M. Sato, Dr. H. Sagawa, and
Dr. Y. Takahashi for useful discussions.

\appendix
\def\thesection{Appendix \Alph{section}}
\def\thesubsection{\Alph{section}-\Roman{subsection}}
\renewcommand{\theequation}{A.\arabic{equation}}
\setcounter{equation}{0}

\section{}
Wave functions including CG coefficients are 
\begin{eqnarray}
& &\phi_0:| J=0 ,{\vec P}=0\rangle  = ({1 \over \sqrt 3}|1/2,1/2;-1\rangle 
-{1 \over \sqrt 6}|1/2,-1/2;-1 \rangle \nonumber\\  
& &~~~~+{1 \over \sqrt 6} |-1/2,1/2;-1\rangle +|-1/2,-1/2;-1\rangle ,
\\
& &\phi_1:|J=1,{\vec P}=0 \rangle=  ({1 \over \sqrt 2}|1/2,1/2;0\rangle -{1 \over
 2}|+1/2,-1/2;-1 \rangle  \nonumber\\
& &~~~~~-{1 \over 2}|-1/2,1/2; 1\rangle,    \\
& &\phi_2:|J=2 {\vec P}=0\rangle=  |1/2,1/2;1\rangle. 
\end{eqnarray}
\section{}
The integration over a semi-infinite region of  satisfying the
causality in   two dimensional variables,
 $x_1(t_1,x_1), x_2(t_2,x_2)$
\begin{eqnarray}
I=\int_{\lambda_1,\lambda_2 \geq 0} d^2x_1 d^2x_2 {\partial \over \partial t_1} {\partial \over \partial
 t_2} f(x_1-x_2)g(x_1+x_2) ,
\end{eqnarray}
where the integrands satisfy
\begin{eqnarray}
g(t_{+},x_{+}=\pm \infty)=0,\ f(t_{-},x_{-}=\pm \infty)=0 \label{b.condition}
\end{eqnarray}
is made with the change of variables. Due to Eq. (\ref{b.condition}),  $I$ would vanish if the integration region were
from $-\infty$ to $+\infty$. We write $I$ with variables $x_{+}={x_1+x_2 \over 2},\ x_{-}={x_1-x_2}$,
and have 
\begin{eqnarray}
& &I= I_1-I_2,\\
& &I_1=\int_{\lambda_{+} \geq 0} d^2x_{-} f(x_{-}) d^2x_{+}  ({\partial^2 \over \partial
 t_{+}^2}g(x_{+}))  \nonumber, \\ 
& &I_2=\int_{\lambda_{+} \geq 0} d^2x_{+}g(x_{+}) d^2x_{-} ({\partial^2 \over \partial
 t_{-}^2} f(x_{-})), \nonumber  
\end{eqnarray}
where $\lambda_{+}=x_{+}^2$. It is noted that   $x_{+}$ is integrated in 
the restricted region but $x_{-}$ is integrated in whole region. 
Thus
\begin{eqnarray}
& &I_1= -\int d^2x_{-} f(x_{-})dt_{+}v^2 ({\partial \over \partial
 x_{+}} g(x_{+}))|_{x_{+}=x_{+,{min}}},\\
& &I_2=0, \nonumber
\end{eqnarray}
where the functional form $g(x_{+})=g(x_{+}-vt_{+})$ was used. 
 $I$ is computed with the slope of $g(x_{+})$ at the boundary $x_{+}=t_{+}$. 
We apply this  method  for computing   four dimensional integrals. 

\section{}
In the case of the low energy region the photon has no effective mass, but is expressed
by the index of refraction very close to unity in dilute gas,
\begin{eqnarray}
n=1+\epsilon.
\end{eqnarray} 
Accordingly, the angular velocity  in $\bar \phi_c(\delta t ) $ is
given by
\begin{eqnarray}
\omega=(1+\epsilon)P_{\gamma}-P_{\gamma}=\epsilon P_{\gamma}.
\end{eqnarray}
$\epsilon$ in air is 
\begin{eqnarray}
\epsilon=0.000292,~~0C, 1~\text {atomspheric~ pressure},
\end{eqnarray}
and $\epsilon$  in
dilute gas  becomes  extremely small  of the order $10^{-14}-10^{-15}$. Consequently  
the integrand in $P^{(d)}$  is proportional to ${2 \over P_{\gamma}\epsilon}$. 

\section{  Notations :MKSA unit}
It is convenient to express  the Lagrangian with MKSA unit to study
quantum phenomena caused by macroscopic electric and magnetic field \cite{Cohen-Tannoudji}. 
The Maxwell equations for 
electric and magnetic fields are 
expressed   with the dielectric constant and magnetic
permeability of the vacuum,  $\epsilon_0, \mu_0$ which are related with
the speed of light, c, 
\begin{eqnarray}
c={1 \over \sqrt {\epsilon_0 \mu_0}}
\end{eqnarray}  
The zeroth component $x_0$ in 4-dimensional coordinates $x_{\mu}$,  is 
\begin{eqnarray}
x_0=ct;\ t=~\text{sec}.
\end{eqnarray}
The Maxwell equations in vacuum are,
\begin{eqnarray}
& &{\vec \nabla} \cdot {\vec E}={1 \over \epsilon_0} \rho(x), \\
& &{\vec \nabla} \cdot {\vec B}=0, \nonumber  \\
& &{\vec \nabla} \times  ({1 \over c}{\vec E})=-{\partial {\vec B }\over \partial x_0 }
 \nonumber , \\
& &{\vec \nabla} \times  {\vec B}=\mu_0 {\vec j(x)}+
 {\partial ({1 \over c}{\vec E}) \over \partial x_0}, \nonumber
\end{eqnarray}
where  the charge density and electric current,
$\rho(x),\ {\vec j}(x)$,
satisfy
\begin{eqnarray}
c{\partial \over \partial x_0} \rho(x)+{\vec {\nabla}}{\vec j}(x)=0.
\end{eqnarray}
Using the vector potential
\begin{eqnarray}
A_{\mu}=(A_0,-{\vec A}\,),
\end{eqnarray}
we write
\begin{eqnarray}
& &{1 \over c}{\vec E}(x)=-{\partial \over \partial x_0}{\vec A}(x)-{\vec {\nabla}}A_0(x),\\
& &{\vec B}={\vec {\nabla}}\times {\vec A}(x) \nonumber,
\end{eqnarray}
and the Lagrangian density of electric and magnetic fields  
\begin{eqnarray}
L_{EM}=-{1 \over 4}{1 \over \mu_0} F_{\mu\nu}F^{\mu\nu}={\epsilon_0 \over
 2} {\vec E}^2-{1 \over
 2 \mu_0 }{\vec B}^2. \label{electromagnetic-MKSA}
\end{eqnarray}
The Lagrangian density of electronic  fields is,
\begin{eqnarray}
& &L_{e}=\bar \psi(x)\left[ \gamma_0 i c \hbar{\partial \over \partial x_0}
 -\gamma_l i c \hbar{\partial \over \partial x_l}-mc^2 \right] \psi  
\label{electron}, 
\end{eqnarray}
and that of QED is
\begin{eqnarray}
L_{QED}=L_{e}+L_{EM}++ecA_0\bar
 \psi \gamma_0 \psi -ecA_l\bar \psi \gamma_l \psi \label{QED-MKSA}.
\end{eqnarray}
Canonical momenta and commutation relations  from
Eqs. $(\ref{electromagnetic-MKSA})$ and $( \ref{QED-MKSA})$ are
\begin{eqnarray}
& &\pi_{\psi}(x)= {\partial \over \partial \dot \psi(x)}L=i\hbar \psi^{\dagger}(x), \\
& &\Pi_l(x)(x)={\partial \over \partial \dot A_i(x)}L=  { 1 \over \mu_0 c^2} E_i \nonumber, \\
& &\{\psi(x_1), \psi^{\dagger}(x_2) \}\delta(t_1-t_2)= \delta({\vec x}_1-{\vec
 x}_2) \delta(t_1-t_2),\\
& &[ A_i(x_1), \dot A_j(x_2)]\delta(t_1-t_2)= i\hbar {1 \over \epsilon_0} \delta_{ij} \delta({\vec x}_1-{\vec
 x}_2)\delta(t_1-t_2), \nonumber
\end{eqnarray}
where the gauge dependent term was ignored in the last equation.
Thus the commutation relations for electron fields do not have $\hbar$,
and those of electromagnetic fields have $\hbar $ and the dielectric
constant $\epsilon_0$.   ${\hbar
\over \epsilon_0}$ shows  the unit size in phase space.  Accordingly 
the number of states per unit area  and the strength
of the light-cone singularity  are  proportional 
to $1/\epsilon_0$.
The fields are expanded with the wave vectors as
\begin{align}
&\psi(x)=\sum_s \int {d{\vec k} \over (2\pi)^{3/2}}({\tilde \omega_0
 \over k^0({\vec k})})^{1/2}(u({ \vec k},s) b({\vec k},s)
 e^{-ikx}+v({\vec k},s) d^{\dagger}({\vec k},s)  e^{ikx}),\\
&A_i(x)=\sum_s \int {d{\vec k} \over ((2\pi)^3 2k_0)^{1/2}}({ \hbar 
 \over  \epsilon_0})^{1/2}(\epsilon_i({ \vec k},s) a({\vec k},s)
 e^{-ikx}+\epsilon^{*}({\vec k},s) a^{\dagger}({\vec k},s)  e^{ikx}),
\end{align}
where
\begin{eqnarray}
& &kx=k_0x_0-{\vec k}{\vec x}, \\
& &k_0({\vec k})=\sqrt{{\vec k}^2+{\tilde \omega_0}^2}, ~~\tilde \omega_0={mc \over
 \hbar}  \nonumber.
\end{eqnarray}
The creation and annihilation operators satisfy
\begin{eqnarray}
& &\{b({\vec k}_1,s_1), b^{\dagger}({\vec k}_2,s_2) \}=\delta({\vec
 k}_1-{\vec k}_2) \delta_{s_1s_2}, \\
& &[a({\vec k}_1,s_1), a^{\dagger}({\vec k}_2,s_2)] =\delta({\vec
 k}_1-{\vec k}_2) \delta_{s_1s_2}. 
\end{eqnarray}
The spinor is normalized as
\begin{eqnarray}
\sum_s u ({\vec k},s) \bar u({\vec k},s)={{\gamma}\cdot k +\tilde \omega_0 \over
 2\tilde \omega_0},
\end{eqnarray}
and the light-cone singularity is 
\begin{eqnarray}
\Delta_{+}(x_1-x_2)=\int {d{\vec k} \over (2\pi)^3 2k_0}
 e^{-ik\cdot(x_1-x_2)}=2i {1 \over 4\pi} \delta(\lambda) \epsilon(\delta
 x^0)+\cdots,
\end{eqnarray}
where the less-singular and regular terms are in $\cdots$.
 
The action 
\begin{eqnarray}
& &S=\int dx (L_{QED}+\bar \psi(x)\gamma_{\mu}(1-\gamma_5)\psi(x)
 J^{\mu}(x)),\\
& &J^{\mu}(x)={G_F \over \sqrt 2}\bar \nu(x) \gamma^{\mu}(1-\gamma_5)
 \nu(x). \nonumber
\end{eqnarray}  
governs the dynamics of electron, photon and neutrino. Integrating
$\psi(x)$ and $\bar \psi(x)$, we have  
$\text{Det}(D+J^{\mu}\gamma_{\mu}(1-\gamma_5))$ and the  effective action 
between neutrino and photon 
\begin{eqnarray}
S_{eff}={e^2 \over 8\pi^2 c \hbar^2} f(0)\int dx{\partial \over
 \partial x_{\nu}}\left[J^{\nu}(x)\epsilon_{\alpha\beta\rho\sigma}{\partial
 \over \partial x_{\alpha}}A^{\beta} {\partial
 \over \partial x_{\rho}}A^{\sigma}\right],\label{action-MKSA}
\end{eqnarray}
where 
\begin{eqnarray}
f(0)={1 \over 2{\tilde \omega_0}^2}.
\end{eqnarray}
The magnetic field and  electric field in the 3-rd direction, and laser
field  expressed by the external fields,
\begin{eqnarray}
F^{\mu\nu}_{ext}=\epsilon^{03\mu\nu} B,\ F^{\mu\nu}_{ext}=\epsilon^{12\mu\nu} {E \over
 c},\ 
 A^{\mu}_{laser}(x)
\end{eqnarray}
are substituted to the action Eq .$(\ref{action-MKSA})$, and we have the actions
\begin{align}
&S_{eff,B}=g_B \epsilon_{03\rho\sigma} \int dx{\partial \over
 \partial x_{\nu}}[\bar \nu(x) \gamma^{\nu}(1-\gamma_5)\nu(x) {\partial
 \over \partial x_{\rho}}A^{\sigma}(x)],\label{action-MKSA1} \\
&S_{eff,E}=g_E \epsilon_{12\rho\sigma} \int dx{\partial \over
 \partial x_{\nu}}[\bar \nu(x) \gamma^{\nu}(1-\gamma_5)\nu(x) {\partial
 \over \partial x_{\rho}}A^{\sigma}(x)],\\
&S_{eff,laser}=g_l \epsilon_{\mu\nu\rho\sigma} \int dx{\partial \over
 \partial x_{\nu}}[\bar \nu(x) \gamma^{\nu}(1-\gamma_5)\nu(x) F^{\mu\nu}_{laser}(x){\partial
 \over \partial x_{\rho}}A^{\sigma}(x)],
\end{align}
where the coupling strengths are
\begin{eqnarray}
& &g_B=  {e^2 \over 8\pi^2 c \hbar^2} f(0) B {G_F \over \sqrt 2},\\
& &g_E= {e^2 \over 8\pi^2 c \hbar^2} f(0) { E \over c}{G_F \over \sqrt 2},\\
& &g_l=  {e^2 \over 8\pi^2 c \hbar^2} f(0) {G_F \over \sqrt 2}.
\end{eqnarray}
The wave vectors are connected with the energy and momentum
\begin{eqnarray}
E({\vec k})=\hbar c k_0({\vec k}),\ {\vec p}=\hbar c {\vec k}.
\end{eqnarray}

\section{Integration formulae}
The integrals 
\begin{align}
& I_0=\int_{\lambda_{+} = 0} d^4x_1 d^4x_2 \rho_s(x_{-}^0)({\partial \over \partial x_{+}^0})^2
 e^{-\chi(x_1)-\chi(x_2)^{*}}\delta((\lambda_{-})^2)
 e^{ip\cdot x_{-}}, \\
&I_i=\int_{\lambda_{+} = 0 } d^4x_1 d^4x_2 \rho_s(x_{-}^0) ({\partial \over \partial x_{+}^i})^2
 e^{-\chi(x_1)-\chi(x_2)^{*}}\delta((\lambda_{-})^2)e^{ip\cdot x_{-}},
 \\  
&\rho_s(x_{-}^0)=1+i{x_1^0-x_2^0 \over E \sigma_{\gamma}} \nonumber,
\end{align}
where the spreading in the transverse direction in Eq. ($\ref{spreading}$)
leads to $\rho_s(x_{-}^0)$ in
the right hand sides.  Changing the variables to $x_{+}$  and $x_{-}$, we have
\begin{eqnarray}
& &\chi(x_1)+\chi(x_2)^{*}=\chi^{(+)}(x_{+})+\chi^{(-)}(x_{-})+\chi^{(+-)}(x_{+},x_{-}), \\
& &\chi^{(+)}(x_{+})={1 \over
 \sigma_{\gamma}}(({x_{+,l}}-v(x_{+,0}))^2+{1 \over
 2\sigma_T(+)}( ({\vec x}_{+})_T^2)\nonumber, \\
& &\chi^{(-)}(x_{-})={1 \over 4\sigma}_{\gamma}
({x_{-,l}}-v(x_{-,0}))^2 +{1 \over
 8 \sigma_T(+)}( ({\vec x}_{-})_T^2) \nonumber, \\
& &\chi^{(+-)}(x_{+},x_{-})={1 \over
 \sigma_T(-)}( ({\vec x}_{+})_T ({\vec x}_{-})_T)\nonumber,
\end{eqnarray}
where the wavepacket sizes in the transverse direction are  
\begin{eqnarray}
& &{1 \over 2\sigma_T(+)}={1 \over 2 \sigma_T(1)}+{1 \over 2
 \sigma_T(2)^{*}}, \\
& &{1 \over 2\sigma_T(-)}={1 \over 2 \sigma_T(1)}-{1 \over 2 \sigma_T(2)^{*}}.
 \nonumber
\end{eqnarray}
The wavepacket expands in the transverse direction and the size
$\sigma_T(+)$  is given by
\begin{eqnarray}
\sigma_T(+)={\sigma_{\gamma} \over 2}-{i \over 4E}(x_{-}^0)+O(x_{-}^2,x_{+}^2).
\end{eqnarray}
The off-diagonal  term $\chi(x_{+},x_{-})$ gives small corrections and
is ignored. We have
\begin{eqnarray}
& &I_0={1 \over \sigma_{\gamma}}I_0(+)I(-),\\
& &I_0(+)=\int_{x_{+}^0=|{\vec x}_{+}|} dx_l^{+} d^2{\vec x}_T^{+}
  e^{-\chi(x_{+})  }{({\vec x}_T^{+})^2 \over
 x_l^{+}}=\int dx_{+}^l\pi (2\sigma_T(+))^2{1 \over x_{+}^l}
 \nonumber, \\
& & I(-)=\int d^4  x_{-} \rho(x_{-}^0)e^{-\chi(x_{-})  }
 \delta(\lambda_{-}^2)e^{ipx_{-}} =\int dx_{-}^0\rho(x_{-}^0){4\sigma_T(+) \over x_{-}^0} e^{i\omega
 x_{-}^0}\nonumber ,
\end{eqnarray}
and 
\begin{eqnarray}
& &I_0=\pi \int_0^T dx_{+}^l {1 \over x_{+}^l} \int_{-x_{+}^l}^{+x_{+}^l} dx_{-}^0 \rho(x_{-}^0){16\sigma_T(+)^3
 \over \sigma_l} { e^{i\omega x_{-}^0} \over x_{-}^0}\nonumber\\
& &=-i 2 \pi^2  ( \sigma_{\gamma}^2 \text {log}(\omega T)+
 { \sigma_{\gamma} \over 4\omega E }). \label{log-constant}
\end{eqnarray}
Eq. $(\ref{log-constant})$ is composed of the $ \log T$ term and
constant. At $\omega  \approx 0$, the latter is important and at a
larger $\omega$, the former is important.  

Similarly
\begin{align}
I_i&=I_i(+)I(-)\\
I_i(+)&=\int_{x_{+}}^i=\pm \sqrt{(x_{+}^0)^2-(x_{+}^{l})^2-(x_{+}^{i})^2}
d x_{+}^0  dx_{+}^l d{\vec x}_{+}^{i'} {x_{+}^i \over \sigma_T(+)} e^{-\chi(x_{+})
}\nonumber\\
&=\pi\int d x_{+}^0
{2 \sigma_T(+) \over  x_{+}^0}\nonumber ,
\end{align}
and
\begin{eqnarray}
& & I_i=\pi \int_0^T dx_{+}^0 {1 \over x_{+}^0} \int_{x_{+}^0}^{x_{+}^0} dx_{-}^0 \rho(x_{-}^0)8\sigma_T(+)^2
 { e^{i\omega x_{-}^0} \over x_{-}^0} \nonumber\\
& &= -i 2\pi^2 (\sigma_{\gamma}^2 \text {log}(\omega T)).
\end{eqnarray}
In high energy regions,  $\omega={m_{\gamma}^2 \over 2E}$, and  
\begin{eqnarray}
& &I_0=-i 2 \pi^2  ( \sigma_{\gamma}^2 \text {log}(\omega T)+
 { \sigma_{\gamma} \over 4m_{\gamma}^2}), \label{high-energy-I}\\
& & I_{T,i}= -i 2\pi^2 (\sigma_{\gamma}^2 \text {log}(\omega T)),\\
& &I_l=I_0.
\end{eqnarray}
In low  energy regions,  $\omega=\epsilon p $, and  
\begin{eqnarray}
& &I_0=-i 2 \pi^2  ( \sigma_{\gamma}^2 \text {log}(\omega T)+
 { \sigma_{\gamma} \over 4 \epsilon p^2 }),  \label{low-energy-I}\\
& & I_i= -i 2\pi^2 (\sigma_{\gamma}^2 \text {log}(\omega T) ),\\
& & I_l=I_0.
\end{eqnarray}


\begin{thebibliography}{}
\bibitem{ishikawa-tobita1} K. Ishikawa and Y. Tobita.
Prog. Theor. Exp. Phys. 073B02, doi:10.1093/ptep/ptt049 (2013).

\bibitem{ishikawa-tobita2} K. Ishikawa and Y. Tobita.
Ann of Phys, \textbf{344}, 118(2014),  doi: 10.1016/j.aop.2014.02.007.

\bibitem{sakurai}
J. J. Sakurai,  Advanced quantum mechanics  (Princeton University
	Press, New Jersey 1979) p.184. 
\bibitem{peierls}
R. Peierls. Surprises in Theoretical Physics (Princeton University
	Press, New Jersey 1979) p.121.
\bibitem{greiner} W. Greiner, QUANTUM MECHANICS An Introduction
(Springer, New York Berlin Heidelberg, 1994), p282.


\bibitem{Dirac} P. A. M. Dirac.  Pro. R. Soc. Lond. A 114, 243 (1927).
\bibitem{Schiff-golden}
L. I. Schiff. 
{\it  Quantum Mechanics}~(McGRAW-Hill
Book COMPANY,Inc.  New York, 1955).   
\bibitem{Goldberger}  M. L. Goldberger and Kenneth  M. Watson, 
{\it Collision Theory} (John Wiley \& Sons, Inc. New York, 1965).

\bibitem{newton}  R. G. Newton, 
\textit{Scattering Theory of Waves and Particles}~(Springer-Verlag, New York, 1982).
\bibitem{taylor}  J. R. Taylor, 
\textit{Scattering Theory: The Quantum Theory of non-relativistic
	Collisions} (Dover  Publications, New York, 2006).

\bibitem{laser-compton} M. Iinuma,et al, Phys.Letters A 346 255-260(2005)
 
\bibitem{LSZ} H. Lehman, K. Symanzik, and W. Zimmermann,  Il~Nuovo~Cimento~(1955-1965)~
\textbf{1},~205 (1955).

\bibitem{Low} F. Low,  Phys. Rev. \textbf{97}, 1392 (1955).



\bibitem{Ishikawa-Shimomura} K. Ishikawa and T. Shimomura,
	Prog. Theor. Phys.  \textbf{ 114}, 1201 (2005) [hep-ph/0508303].
\bibitem{Ishikawa-Tobita-ptp} K. Ishikawa and Y. Tobita.
	Prog. Theor. Phys.  \textbf{ 122}, 1111 (2009) [arXiv:0906.3938[quant-ph]]. 

\bibitem{Ishikawa-Tobita} K. Ishikawa and Y. Tobita, 
AIP Conf. proc. \textbf{1016}, 329(2008);
arXiv:0801.3124 [hep-ph].

 \bibitem{Ishikawa-Tobita-prl} K.~Ishikawa and Y.~Tobita, arXiv:1106.4968[hep-ph].

\bibitem{Kayser} B.~Kayser.~Phys.~Rev. \textbf{D24},~110 (1981).

\bibitem{Giunti} C.~Giunti, C.~W.~Kim and U.~W.~Lee.  Phys. Rev. \textbf{D44}, 3635 (1991).
\bibitem{Nussinov} S.~Nussinov. Phys. Lett. \textbf{B63}, 201 (1976).

\bibitem{Kiers} K.~Kiers, S.~Nussinov and N.~Weiss. 
	Phys. Rev. \textbf{D53}, 537 (1996) [hep-ph/9506271].
\bibitem{Stodolsky} L.~Stodolsky.  Phys. Rev. \textbf{D58}, 036006 (1998) [hep-ph/9802387].

\bibitem{Lipkin} H.~J.~Lipkin.~Phys.~Lett.~\textbf{B642},~366 (2006) [hep-ph/0505141].

\bibitem{Akhmedov} E. K.  Akhmedov.~JHEP. \textbf{0709}, 116 (2007) [arXiv:0706.1216 [hep-ph]].

\bibitem{Asahara} A.~Asahara, K.~Ishikawa, T.~Shimomura, and T.~Yabuki,
Prog. Theor. Phys. \textbf{113}, 385 (2005) [hep-ph/0406141]; T.~Yabuki and K.~Ishikawa.
Prog. Theor. Phys. \textbf{108}, 347 (2002).


\bibitem{Anderson} H.~L.~Anderson et al.  Phys. Rev. \textbf{119},
	2050 (1960).
\bibitem{homma-tajima1} K. Homma, D. Habs, and T. Tajima, Appl. Phys. \textbf{B106}, 229(2012).

\bibitem{homma-tajima2}T. Tajima, and K. Homma, Int. J. Mod. Phys\textbf{A27},1230027(2012).

\bibitem{tajima-shibata}C. H. Lai, Ph.D dissertation (Univeristy of
	Texas,Austin,1994); T. Tajima and K. Shibata, ''Plasma
	Astrophysics'' (Addison-Wesley,Reading,1997) p.451.

\bibitem{landau} L.~Landau. Sov. Phys. Doclady \textbf{60}, 207(1948).

\bibitem{yang} C. N. Yang.  Phys. Rev. \textbf{77},
	242(1950).

\bibitem{fukuda-miyamoto} H. Fukuda and Y. Miyamoto, Prog. Theor. Phys.\textbf{4},49(1949).

\bibitem{steinberger} J. Steinberger, Phys. Rev. \textbf{76}, 1180(1949).

\bibitem{rosenberg} L. Rosenberg. Phys. Rev. \textbf{129}, 2786(1963).

\bibitem{adler} S. ~L. Adler. Phys. Rev. \textbf{177}, 2426(1969).

\bibitem{liu} J. Liu, Phys. Rev. D\textbf{44}, 2879(1991). 

\bibitem{feynman} R. P. Feynman, Phys. Rev. \textbf{76}, 749(1949).

\bibitem{aleskseev} A. I. Alelseev,
	Zh. Eksp. Teor. Fiz. \textbf{34},1195(1958)
 [Sov Phys. JETP 7,826(1958)].

\bibitem{tumanov} K. A. Tumanov,
	Zh. Eksp. Teor. Fiz. \textbf{25}, 385(1953).
\bibitem{tajima-dawson} T. Tajima and J. M. Dawson,
 Phys. Rev. Lett. \textbf{43}, 267(1979). 

\bibitem{barbieri} R. Barbieri, R. Gatto, and R. Kogerler, Phys. Lett. \textbf{60B}, 183(1976).


\bibitem{kwong} W. Kwong,P. B. Mackenzie, R. Rosenfeld, and J. L. Rosner,
 Phys. Rev. D\textbf{37},3210(1988). 
\bibitem{li} Z. P. Li, F.E.Close, and T.Barns,
	Phys. Rev. D\textbf{43}, 2161(1991).  
\bibitem{bodwin} G.T. Bodwin, E. Braaten and G. P. Lepage, 
	Phys. Rev. D\textbf{51}, 1125(1995).  
\bibitem{huang} Han-Wen Huang and Kuang-Ta Chao,
 Phys. Rev. D\textbf{54}, 6850(1996). 





\bibitem{crystal-ball} J. E. Gaiser, et al,
	Phys. Rev. D\textbf{34}, 711(1986).
 
\bibitem{cleo} S. B. Athar, et al,
	Phys. Rev. D\textbf{70}, 112002(2004).

\bibitem{bes} M. Ablikin, et al,
	Phys. Rev. D\textbf{71}, 092002(2005).

\bibitem{particle-data}
  J. Beringer {\it et al.}  [Particle Data Group],
  J.\ Phys.Rev  \textbf{D86}, 010001(2012).

\bibitem{gell-mann} M. Gell-Mann. Phys. Rev. Lett. \textbf{6}, 70(1961).

\bibitem{axion} Y. Inoue, et. al, Phys. Lett. \textbf{B668}, 93(2008). 


 \bibitem{relic-neutrino} J.Arafune and G.Takeda, ``Total Reflection of
	 Relic Neutrinos from Material Targes'', University of
	 Tokyo, ICEPP Report, ut-icepp 08-02, and  private communication.

 \bibitem{ice-cube}M. G. Aartsen et al. (IceCube Collaboration): Observation of High-Energy Astrophysical
Neutrinos in Three Years of IceCube Data, 2014, arXiv:1405.5303 [astro-ph.HE].
\bibitem{auger}
P. Abreu et al. (Pierre Auger Collaboration), Adv. High
Energy Phys. 2013, 708680 (2013), arXiv:1304.1630
[astro-ph.HE].
\bibitem{telescope-array}
G. I. Rubtsov et al. (Telescope Array Collaboration),
 J. Phys. Conf. Series \textbf{409} (2013), 012087.

\bibitem{upper-lightening} R. C. Franz, R. J. Nemzek, and
	J. R. Winckler, Science \textbf{249} 48 (1990); G. J. Smith, et.al.,
	Science,textbf{264},1313(1994).        

 \bibitem{gamma-cumulonimbus } T.Torii, M.Takeishi, and T. Hosono,
	 J.Geophys.Res.\textbf{107},4324(2002); H.Tsuchiya, et. al, Phys.Rev.Letters. \textbf{99}165002(2007)

\bibitem{promordial-magnetic-field} T.Tajima, S. Cable, K.Shibata, and
	R. M. Kulsrud, Ap. J. \textbf{390}, 309 (1992).

\bibitem{Cohen-Tannoudji}
Claude Cohen-Tannoudji,~Jacques Dupont-Roc, and Gilbert Grynberg.
{\it Photons and atoms: Introduction to Quantum Elctrodynamics}~(John Wiley \&
	Sons, Inc. New York, 1989).


\end{thebibliography}
\end{document}